\documentclass{aastex63}
\def\simgr{\,\hbox{\hbox{$ > $}\kern -0.8em \lower 1.0ex\hbox{$\sim$}}\,}
\def\simle{\,\hbox{\hbox{$ < $}\kern -0.8em \lower 1.0ex\hbox{$\sim$}}\,}

\newcommand{\numberOfTargets}{30}

\shortauthors{THORSTENSEN}
\shorttitle{Short-Period Cataclysmics}

\begin{document}
\title{Spectroscopic Studies of \numberOfTargets\ Short-period 
Cataclysmic Variable Stars,\\ and Remarks on the Evolution and
Population of Similar Objects
}

\author[0000-0002-4964-4144]{John R. Thorstensen}

\affil{Department of Physics and Astronomy\\
6127 Wilder Laboratory, Dartmouth College\\
Hanover, NH 03755-3528}

\begin{abstract}
We present spectroscopy and orbital periods $P_{\rm orb}$ 
for \numberOfTargets\
apparently non-magnetic cataclysmic binaries with periods 
below $\sim 3$ hours, nearly all of which are dwarf novae,
mostly of the SU Ursae Majoris subclass.
We then turn to the evidence supporting the prediction that 
short-period dwarf novae evolve toward
longer periods after passing through a minimum period -- the 
`period bounce' phenomenon.  Plotting data from the literature
reveals that for superhump period excess
$\epsilon = (P_{\rm sh} - P_{\rm orb}) / P_{\rm orb}$ 
below $\sim 0.015$, the period appears to 
increase with decreasing $\epsilon$, agreeing 
at least qualitatively with the predicted behavior.  
Next, motivated by the long ($\sim$ decadal) outburst
intervals of the WZ Sagittae subclass of short-period
dwarf novae, we ask whether there could be a sizable 
population of `lurkers' -- systems that resemble dwarf novae at 
minimum light, but which do not outburst over 
accessible timescales (or at all), and therefore
do not draw attention to themselves.  By examining the 
outburst history of the Sloan Digital Sky Survey sample of CVs,
which were selected by color and not by outburst, we find that 
a large majority of the color-selected dwarf-nova-like 
objects have been observed to outburst, and conclude that 
`lurkers', if they exist, are a relatively minor part of the 
CV population.

\end{abstract}


\section{Introduction}

Cataclysmic variable stars (CVs) are close binary systems consisting
of a degenerate dwarf primary and a more extended star that 
transfers mass onto the degenerate component through Roche
lobe overflow.  Most CVs appear as dwarf novae, which undergo
outbursts of 2-10 magnitudes from a quiescent state.  The
outburst interval varies widely, from nearly continuous to
many years.  The cause of the outburst is thought to be an
accretion disk instability that occurs when material
accumulating in the disk reaches a critical optical depth 
\citep{mmh81,smak84}.

The SU Ursae Majoris stars \citep{patt79,vogt80} are a subclass of 
dwarf novae that show {\it superoutbursts}.  These are distinguished 
from normal outbursts by their higher amplitudes and longer duration.
Within (typically) several days after the onset
of a superoutburst, {\it superhumps} develop.  Superhumps are 
photometric oscillations at a nearly constant period, $P_{\rm sh}$ 
that is almost always slightly {\it longer} than the orbital 
period $P_{\rm orb}$.  Nearly all SU UMa stars have $P_{\rm orb}$ 
shortward of the $\sim 2$-3 hr `gap' in the $P_{\rm orb}$ distribution.
The generally accepted explanation for superhumps 
(see, e.g., \citet{whitehurst91})
begins with the fact that in these short-period systems, the 
mass ratio $q = M_2 / M_{\rm WD}$ is low, which is a consequence
of the condition that the secondary fills its Roche critical 
lobe \citep{faulkner72,patt84,knigge06}.
This makes the (mostly empty) Roche lobe around the white dwarf
relatively large, making it possible for the  
outbursting disk to expand enough that 
the disk's edge reaches a 3:1 resonance with the 
orbit\footnote{At short periods, the 3:1 resonance is also
closer to the white dwarf, since the period of the 
resonance is correspondingly short.}.
After the disk expands to the resonant radius, 
it becomes elongated and precesses slowly
in the inertial frame.  The tidal stresses raised by
the secondary in the disk are greatest when the secondary aligns
with the disk's long axis.  Because of the precession, 
the disk brightens at a frequency
slightly lower than the orbital frequency, since 
each successive passage of the secondary through the disk's long 
axis occurs a little later in orbital phase.  

Some authors define a WZ Sagittae subclass of the 
SU UMa stars, which (1) undergo only superoutbursts, with 
no `normal' ones, and (2) typically outburst
at intervals of a decade or more.  WZ Sge stars tend to 
have very short $P_{\rm orb}$s, typically near 75-85 min.  

The fractional superhump period excess, 
\begin{equation}
\epsilon = {P_{\rm sh} - P_{\rm orb} \over P_{\rm orb}},
\end{equation}
is thought to be related to the mass ratio $q$. 
This is physically plausible since the secondary's gravity
drives the disk precession.  \citet{patt05} used known
mass ratios (derived largely from eclipse studies) to calibrate 
the relationship between $q$ and $\epsilon$.  As noted
earlier, the mass
ratio of a CV is strongly correlated with $P_{\rm orb}$,
because the secondaries of longer-period CVs must fill 
correspondingly larger Roche lobes.  Consequently, the 
underlying $q(\epsilon)$ relation creates the observed
relationship between $\epsilon$ and $P_{\rm orb}$.  
In non-eclipsing systems, $P_{\rm orb}$ can be difficult
to establish; by contrast, superoutbursting stars
are bright enough, and the superhumps distinct enough, 
that skilled amateur astronomers with small telescopes 
and CCD photometers
routinely measure superhump periods.  
Assuming that $\epsilon$ is typical, one can 
infer $P_{\rm orb}$ from $P_{\rm sh}$ to within about one percent.  

The mass ratio, for which $\epsilon$ serves as a proxy,
is a key parameter in CV evolution.  In 
nearly all CVs, $q < 1$, and the mass transfer is stable in the
sense that, when mass is lost from the secondary, its radius
should decrease more quickly than the size of its Roche lobe,
shutting off mass transfer.  The evolution of
CVs is therefore thought to be driven by the loss of orbital 
angular momentum, probably through magnetic braking of the
co-rotating secondary at long periods, with gravitational
radiation dominating at the shortest periods.  There is 
one startling prediction -- around $P_{\rm orb} \sim 70$ min,
the secondary becomes degenerate and its radius 
begins to {\it increase} with decreasing mass.  This leads
to a gradual outspiral of the orbit following the period 
minimum. CVs in this stage are referred to as {\it post-bounce
CVs}, though the reversal of the period evolution should
be much more gradual than the phrase suggests. If post-bounce 
systems exist, they should be distinguished by very low mass ratios,
and hence anomalously small $\epsilon$ for their $P_{\rm orb}$.  
\citet{pattmurmurs} presents a lucid and wide-ranging 
discussion of this phenomenon.  

Dwarf nova outbursts are not the only channel through which
CVs are discovered.  Others are found from their X-ray emission
(see \citealt{halpern2} for a recent example), peculiar
colors (typically ultraviolet excesses; see, e.g., the
series of Sloan Digital Sky Survey (SDSS) CV papers ending with \citealt{szkodyviii}), 
or in spectral surveys (e.g., \citealt{hqs06}, \citealt{witham07,witham08}).  
Many of the objects found through channels other than 
outburst searches are `novalike variables', a catchall class that 
has come to mean `CVs that are not dwarf novae'.  Some appear to
be in extended outburst states, as if their disks are 
in a stable high state; these include the UX UMa stars and
the SW Sex stars.  Many others have highly magnetized 
white dwarfs, and appear as AM Her stars (also called {\it polars})
or DQ Her stars ({\it intermediate polars}).

The space density of CVs constrains evolutionary and
population-synthesis models.  However, because of the 
large variety of discovery channels, it is difficult to 
estimate the completeness of the known sample of CVs.  
The majority of CVs were found because they are
outbursting dwarf novae.  However, the outburst 
interval introduces a `wild card' into the completeness
estimate; the outburst intervals of WZ Sge stars can 
exceed a decade, and there is no obvious reason why
they could not be much longer.  The WZ Sge stars all
have orbital periods near the $\sim$75 minute minimum;
if there is a substantial population of ``lurkers'' --
dwarf novae with extremely long recurrence intervals --
we expect them to be found among the shortest orbital
periods, which we focus on here.  We will return to this
question in Section \ref{subsec: lurkers}.

Here we present spectroscopic periods for
\numberOfTargets\ CVs that prove to have periods shortward of,
or near, the 2-3 hr gap.  One motivation is to refine and strengthen the 
$\epsilon$-$P_{\rm orb}$ relation.  Many CVs in outburst
are quickly found to be superhumping, and $P_{\rm sh}$ is
used to estimate $P_{\rm orb}$ (see e.g. \citealt{unveils});
this practice depends on the $\epsilon$-$P_{\rm orb}$ relation, 
but and obviously cannot be used to improve that relation
without an independent determination of $P_{\rm orb}$.  
An independent $P_{\rm orb}$ is 
also needed to reveal post-bounce systems, if they exist.  Leaving 
aside the $\epsilon$-$P_{\rm orb}$ relation,
orbital periods of non-outbursting systems are needed to
see if they be ``lurking'' WZ Sge stars.
Finally, the good signal-to-noise resulting from the 
long cumulative exposures required for spectroscopic periods 
sometimes reveals spectral anomalies.

\section{Observations and Reductions}

All the observations reported are from the MDM
Observatory on Kitt Peak, Arizona, and nearly all are
spectra.  

Most are from the 
`modspec' spectrograph, mounted either on the Hiltner 2.4m telescope 
or the McGraw-Hill 1.3m telescope.
For nearly all modspec observations,
the detector was a SITe 2048$^2$ CCD giving 2 \AA\ per 
24 $\mu$m pixel.
and a useful range from 4310 to 7500 \AA, with severe vignetting
toward the ends of the range.  A few observations were taken with a
1024$^2$ SITe chip with the same pixel spacing, set up to
cover from 4600 to 6800 \AA.  For wavelength calibration we
used spectra of Hg, Ne, and Xe comparison lamps taken in twilight
to establish the shape of the pixel-wavelength relation, and then 
adjusted for instrument flexure during the night using the
strong $\lambda$5577 night-sky line.  We also observed flux
standard stars, but our average fluxes suffer from estimated 
uncertainties of at least 20 per cent due to cloudy intervals and
uncalibrated losses at the slit, which had a projected width of
$1.''1$ at the 2.4 m and $2.''0$ at the 1.3 m.  
Spectra taken with the modspec often show unphysical variations
in the continuum shapes, which we do not understand, but these 
do tend to average out over large numbers of exposures.  

Our most recent observations are from the Ohio State Multi-Object
Spectrometer (OSMOS; \citealt{martini}), mounted on the 2.4m and 
configured to give a spectrum from 3975 to 6870 \AA , with a 
dispersion of 0.7 \AA\ pixel$^{-1}$
and a FWHM resolution of $\sim 4$ \AA.  The wavelength calibration 
of the OSMOS is not as stable as that of the modspec, so we maintained an 
accurate wavelength scale by taking comparison lamps 
before and/or after our object spectra, and using these to 
both shift and `stretch' calibrations derived from lamp spectra taken
at the zenith. Target acquisition with OSMOS 
requires at least one direct image.  We took
those (generally 20 s exposures) with a Sloan $g$ filter, 
and used the PAN-STARRS PSF magnitudes of many stars in 
the field to derive the magnitude zero point and hence
the target's $g$ magnitude just before we took our spectra.
As with the modspec data, we derived a spectrophotometric 
calibation from standard stars; the flux-calibrated continua 
with OSMOS were usually more physically plausible than with modspec.

The short-period CVs studied here undergo ten or more orbital
cycles per day.  Our observations -- from a single site, at 
night -- are necessarily taken on a roughly 24-hour cycle.
This creates aliasing in the period determination -- if the
sampling were strictly periodic, the number of cycles occurring
during the day would be unconstrained.
To ameliorate this, we break the 24-hour sampling periodicity
by observing at large hour angles, where atmospheric dispersion 
\citep{filippenko82} can be problematic.
By default, the spectrograph slit is oriented north-south, which 
is on the parallactic angle at the meridian.
For observations away from the meridian, we compute $|\tan Z \sin x|$, where 
$Z$ is the zenith distance and $x$ is the angle between the position angle of the 
slit and the parallactic angle; any dispersion effect will be 
proportional to this {\it cross-slit refraction
factor}.  For our spectral coverage, slit width, and typical seeing,
we reckon that dispersion losses should be acceptable when the
cross-slit factor is substantially less than 1.  The observation planning
software {\it JSkyCalc} (written by the author) includes a tool for
computing this factor as a function of time; when it 
shows the cross-slit refraction factor growing too large, we
rotate the instrument to an appropriate angle, except for 
at the 1.3 m, where rotation is less convenient and the
wider projected slit ameliorates the problem.

We reduced the data mostly with python-language scripts calling 
IRAF tasks through Pyraf.  For extracting 1-dimensional spectra from the
two-dimensional images, we mostly used a C-language implementation of
the algorithm described by \citet{horne}, while for the
most recent OSMOS data we used a python implementation of 
Horne's algorithm.  The H$\alpha$
emission line invariably gave the most useful radial 
velocities; to measure these, we used a convolution algorithm
described by \citet{sy80}.  For broader lines, we used dual-gaussian
convolution functions that emphasized the steep sides of the line
profile, and for narrow lines we used the 
derivative of a gaussian optimized for the line's width. 
To estimate velocity uncertainties, we took the 
uncertainty in each pixel (computed from the gain, read noise, and 
background), and propagated them forward to compute the error in 
location of the convolution zero-crossing. This is a lower limit, since it
ignores systematic offsets caused by line profile 
variations and other effects.

To search the velocity time series for periods, we used a
`residual-gram' algorithm that fits least-squares sinusoids at
each of a dense grid of fixed frequencies; the figure of merit
is $1/\chi^2$, where $\chi^2$ is the mean of the squared residuals
scaled to their estimated errors.  Well-sampled data from 
multiple nights show alias frequencies separated from the 
best frequency by 1 cycle d$^{-1}$ because of the 
cycle count issue discussed earlier. To assess the reliability of the
alias choice, we used the Monte Carlo test described by \citet{tf85}.

Once a period was determined, we fit the time series with
sinusoids 
\begin{equation}
v(t) = \gamma + K \sin[2 \pi(t - T_0) / P],
\end{equation}
where the epoch $T_0$ was chosen to be near the mean epoch
of the observations (to minimize the correlation between
$T_0$ and $P$), but within one cycle of an observation.
To estimate parameter uncertainties, we used the prescription of 
\citet{cash79}, which in practice amounts to perturbing each
parameter until $\chi^2$ is roughly $1 + 1/N$ of its
minimum value, where $N$ is the number of data points.
The Monte Carlo tests \citep{tf85} generate parameter 
uncertainties as a byproduct; these agree well with those
calculated from the other procedure.

In many cases we have observations from two or more widely
separated observing runs, which are fit well by large numbers
of possible precise periods corresponding to integer numbers of
cycles between runs.  If rough periods could be estimated from the 
isolated data of more than one run, we estimated the period
by fitting each run's velocities individually and
taking a weighted average. 

\section{Results}

Below, we discuss the stars individually in order of right ascension; 
Table~\ref{tab:star_info} gives names and accurate coordinates. 
Table~\ref{tab:velocities} lists radial velocities derived from the 
individual exposures, and serves as a record of when the stars
were observed.  Figures~\ref{fig:montage1} -- \ref{fig:montage3} 
and \ref{fig:montage4} -- \ref{fig:montage8} 
show mean spectra, periodograms, and folded velocity data
for each star.  Table~\ref{tab:parameters} gives parameters
of sinusoidal fits to the velocities.

\begin{deluxetable}{llrrl}
\tablewidth{0pt}
\tablecolumns{5}
\tablecaption{\label{tab:star_info} List of Objects}
\tablehead{
\colhead{Name} &
\colhead{$\alpha_{\rm ICRS}$} &
\colhead{$\delta_{\rm ICRS}$} & 
\colhead{$G$} &
\colhead{$1/\pi_{\rm DR2}$} \\
\colhead{} &
\colhead{[h:m:s]} &
\colhead{[d:m:s]} &
\colhead{} &
\colhead{[pc]} \\
}
\startdata
FL Psc &  00:25:11.038  & +12:17:11.81 & 17.53 & 153.5(+3.3,$-$3.1) \\
1RXS J012750.5+380830 J0127+3808  & 01:27:50.595  & +38:08:11.83 & 17.24 & 366(+34,$-29$) \\
CRTS CSS121120 J020633+205707 & 02:06:33.463 & +20:57:07.19 & 18.25 & 490(+60,$-50$) \\
WY Tri   & 02:25:00.476  & +32:59:55.55 & 17.63 & 490(+35,$-$30) \\
BB Ari   & 02:44:57.797  & +27:31:09.14 & 18.45 & 353(+25,$-$22)\\
SDSS J032015.29+441059.2  & 03:20:15.287  & +44:10:59.20 & 18.87 & 473(+62,$-$49) \\
MASTER OT J034045.31+471632.2 & 03:40:45.292 & +47:16:31.56 & 18.71 & 680(+182,$-$119) \\
V1024 Per  & 04:02:39.048  & +42:50:45.82 & 17.06 & 251.0(+4.7,$-$4.5) \\
V1389 Tau  & 04:06:59.822  & +00:52:43.81 & 18.40 & 561(+71,$-$57) \\
Gaia 19emm & 04:34:36.596 & +18:02:44.79 & 17.40 & 439(+32,$-28$) \\
V1208 Tau  & 04:59:44.043  & +19:26:22.72 & 18.62 & 555(+137,$-$92) \\
ASAS-SN 15pq & 05:21:15.867 & +25:13:32.15 & 16.94 & $254 \pm 8$ \\
1RXS J05573+685  & 05:57:18.463  & +68:32:26.78 & 18.29 & 678(+82,$-$66) \\
SDSS J0751+10 & 07:51:16.993 & +10:00:16.26 & 18.43 & 480(+50,$-$40) \\
SBS 0755+600  & 07:59:26.372  & +59:53:51.06 & 18.31 & 435(+43,$-$36) \\
TT Boo  & 14:57:44.746  & +40:43:40.50 & 18.95 & 679(+105,$-$80) \\
QZ Lib  & 15:36:16.018  & $-$08:39:08.60 & 18.88 & 187(+12,$-$11) \\
CSS170517:155156+1453334 & 15:51:55.625 & +14:53:33.08 & 18.43 & 1400(+1200,$-$450) \\
SDSS J155720.75+180720.2  & 15:57:20.764  & +18:07:20.19 & 18.76 & 847(+258,$-$166) \\
IX Dra  & 18:12:31.451  & +67:04:45.61 & 17.40 & 790(+39,$-$36) \\
ASAS-SN 13at & 18:21:22.503 & +61:48:55.16 & 19.40 & 870(+280,$-$170) \\
KIC 11390659 & 18:58:30.923 & +49:14:32.65 & 16.37 & $259 \pm 3$ \\
RX J1946.2-0444  & 19:46:16.357  & $-$04:44:56.86 & 17.44 & 365(+14,$-$13) \\
V1316 Cyg  & 20:12:13.648  & +42:45:50.95 & 18.07 & 661(+65,$-$55) \\
ASAS-SN 14ds & 20:44:55.871 & $-$11:51:52.55 & 16.82 & 393(+15,$-$14) \\
V444 Peg  & 21:37:01.839  & +07:14:45.69 & 18.64 & 405(+43,$-$35) \\
ASAS-SN 14cl & 21:54:57.695 & +26:41:12.95 & 18.19 & 261(+14,$-$12) \\
V368 Peg  & 22:58:43.478  & +11:09:11.92 & 19.06 & 537(+138,$-$91) \\
IPHAS 2305 & 23:05:38.375 & +65:21:58.63 & 18.79 & 617(+103,$-$77) \\
NSV14652  & 23:38:48.669  & +28:19:54.82 & 18.38 & 627(+102,$-$77) \\
\enddata
\tablecomments{Positions, mean $G$ magnitudes, and distances 
from the GAIA Data Release 2 (DR2; \citealt{GaiaPaper1,GaiaPaper2}).  
Positions are referred to the 
ICRS (essentially the reference frame for J2000), and the catalog 
epoch (for proper motion corrections) is 2015.  The distances and
their error bars are the inverse of the DR2 parallax $\pi_{\rm DR2}$,
and do not include any corrections for possible systematic errors.}
\end{deluxetable}

\subsection{FL Psc (= ASAS 002511+1217.2)}

This object was object was discovered in superoutburst by the
All Sky Automated Survey (ASAS; \citealt{asas}) on 2004 Sep.~11 UT, 
at $V = 10.49$ \citep{price04}.  \citet{templeton06} present a study 
of the superoutburst, and
find superhumps with a period of 0.05687(1) day during the 
interval from 5 to 18 days after the discovery.  \citet{kato1},
find a significantly larger value, $0.057093(12)$, from an
analysis of a more complete data set and by restricting their
analysis to the `plateau' stage of the outburst.
After this, the source
faded to magnitude $\sim 15$ for a few days before undergoing
a single brief `echo' outburst around day 25; during this
fainter interval a photometric period near 0.05666(3) day was detected,
which \citet{templeton06} suggested was probably not orbital because it was
too close to the superhump period.  A brief series of spectra
during the decline showed strong, double-peaked Balmer lines,
modulated in velocity at about 82 minutes; the shortness of the 
time series precluded a precise determination of $P_{\rm orb}$.

FL Psc has apparently remained quiescent since its 2004 outburst,
though an outburst could have been missed due to its low ecliptic
latitude.  Its CRTS DR2 light curve covers from 2004 to 2014, 
and shows it near $\sim 17.3$, with variations of typically a few 
tenths of a magnitude.  

Most of our spectra were taken 2004 Nov 18 -- 21 UT, starting 68 days after the
initial detection, after the source had returned to quiesence
\citep{templeton06}.  The mean spectrum (Fig.~\ref{fig:montage1})
shows a blue continuum with double-peaked Balmer emission.
At the suggestion of T. Kato, we obtained more spectra in 
2018 November with a 6-day time span to improve the precision
of $P_{\rm orb}$ to 0.05604(9) d.  The number of cycles
that elapsed in the 14-year interval between observing runs is 
unknown.  \citet{kato1} found a candidate $P_{\rm orb} = 0.056540(3)$ d
from photometry taken late in the superoutburst.  This is 
significantly longer than the spectroscopic period
we find here, 0.05604(9) d; our spectroscopic period gives 
a significantly larger superhump period excess than the 
photometric period.


\subsection{1RXS J012750.5+380830}

\citet{hu98} discovered this CV (hereafter referred to as
RX J0127+38) as the optical counterpart of a ROSAT X-ray source;
it is listed in the 
\citet{downes} catalog.  The CRTS2 light curve includes 291 points 
between 2005 and 2014, and shows irregular variations by a few tenths
of a magnitude around a typical magnitude of 16.9, but no outbursts.
An outburst to $V = 14.8$ is reported in vsnet-alert 19909
\footnote{Vsnet-alert is an email service to alert
observers to interesting targets; an explanatory
page is found at
{\tt http://ooruri.kusastro.kyoto-u.ac.jp/mailman/listinfo/vsnet-alert},
which links to monthly archives of past alerts.  Many archived
alerts were posted by T. Kato of Kyoto University. Archived
alerts are challenging to reference in the traditional format,
since they are not indexed at ADS, SIMBAD, or other sites.
We adopt the practice of citing these by number, and
where possible mentioning the observers who contributed.
We only cite vsnet-alert when it provides information we
have been unable to find elsewhere.  Section~\ref{subsec: lurkers}
gives more detail about how we access vsnet-alert messages.}.  

\begin{figure}
\includegraphics[height=23 cm,trim = 2.2cm 1cm 2cm 2.8cm,clip=true]{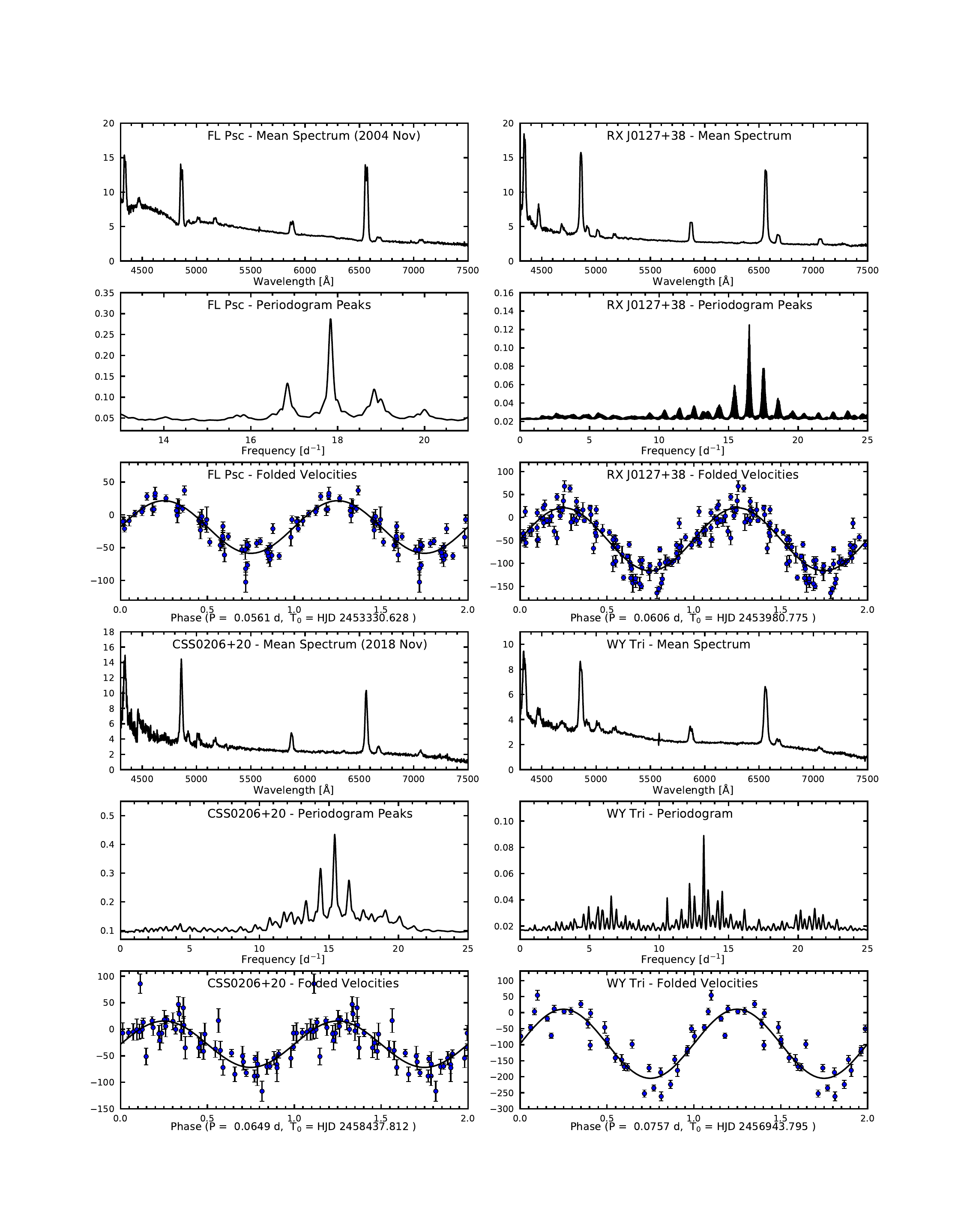}
\caption{{\it Caption on next page.}}
\label{fig:montage1}
\end{figure}

Our mean spectrum (Fig.~\ref{fig:montage1}) has 
good signal-to-noise and shows relatively weak features
such as the iron feature near $\lambda 5169$ and \ion{He}{2} $\lambda 4686$.  
The Balmer lines are strong and just barely double-peaked.

\addtocounter{figure}{-1}
\begin{figure}
\caption{Average spectra, periodograms, and folded velocity
curves for FL Psc, RX J0127, CSS 0206+20, and WY Tri.
The vertical scales, unlabeled to save space, are (1)
for the spectra, $f_\lambda$ in units of $10^{-16}$ erg
s$^{-1}$ cm$^{-2}$ \AA$^{-1}$; (2) for the periodograms,
$1 / \chi^2$ (dimensionless); and (3) for the radial velocity
curves, barycentric radial velocity in km s$^{-1}$.  
In cases where velocities are from more than one observing run, 
the periodogram is labeled with the word ``peaks'', because the 
curve shown is formed by joining local maxima in the 
full periodogram with straight lines. This suppresses fine-scale
ringing due to the unknown number of cycle counts between runs. 
The folded velocity curves all show the same data plotted 
through two cycles for continuity, and the best-fit sinusoid (see 
Table \ref{tab:parameters}) is also plotted.  The velocities shown
are H$\alpha$ emission velocities.  
}
\end{figure}


\subsection{CRTS CSS121120 J020633+205707} 

This object, hereafter referred to as CSS 0206+20, is 
one of many dwarf novae discovered in the Catalina Real Time
Transient Survey (CRTS; \citealt{drakecrtts,
breedt14, drake14}).
A possible superoutburst was noted by Kato in vsnet-alert 20577, but 
superhumps have apparently not been detected.  (See Fig.~\ref{fig:montage1}.)


\subsection{WY Tri}

WY Tri was discovered by \citet{meinunger86}. 
\citet{liuhu00} obtained a spectrum showing the 
broad emission lines typical of a CV at minimum light, and \citet{vanmunster01}
found superhumps with a period of 0.07847(2) d, establishing 
the object as an SU UMa star. (See Fig.~\ref{fig:montage1}.)


\subsection{BB Ari}

BB Ari was discovered by \citet{ross27}. 
Its CRTS2 light curve has 268 detections between
2005 and 2014, and shows variation mostly between
17.8 and 18.8 mag in quiescence, with only a single eruption 
to 14.3 in late 2012.  ASAS-SN \citep{shappeecurtain} detected the object
at 13.9 in 2013 August -- a brightening missed by 
CRTS2 -- and a superoutburst followed. \citet{kato6}
analyzed data from this; they give superhump periods
$P_1 = 0.072544(97)$ and $P_2 = 0.072135(46)$ day.

Our spectra (Fig.~\ref{fig:montage2}) were taken on four observing runs.
The entire velocity time series is fit best by 
$P = 0.0702472(1)$ d, but this level of precision
assumes that the cycle count between 
observing runs is unambiguous.  Monte Carlo tests suggest 
that the cycle-count choice can be made with $\sim 90$ per 
cent confidence.  The next two, less-likely candidate periods 
are 0.0701099 and 0.0699820 d; the first of these differs
from the best period by one cycle per 36 days.

\subsection{SDSSJ032015.29+441059.2} 

This object (SDSS J0320+44 herafter), was identified as a 
CV by \citet{wils10}, who searched for
dwarf novae by data mining and cross-comparing several catalogs.
It has $g = 18.77$ in the SDSS, and they found outbursts to
14.7 mag on two occasions in archival catalogs.  Its sky 
position is not covered in the CRTS2.  Several outbursts
have been reported on vsnet-alert since its discovery, but 
superhumps have apparently not been seen.
(See Fig.~\ref{fig:montage2}.)

\subsection{MASTER OTJ034045.31+471632.2} %

This object (Fig.~\ref{fig:montage2}) was discovered by the MASTER survey at
15.6 mag on 2013 December 26 \citep{atel5698}; we will refer to it 
as OT J0340+47.
The authors note a previous outburst on a 1993 Sky Survey plate, and that
the object's location is not covered by CRTS.  Photometric monitoring
by \citet{hardy17} did not find an eclipse.
A superoutburst occurred in early 2019, and Vanmunster
reported superhumps with an amplitude of 0.21 mag and a period
$P_{\rm sh} = 0.0803(6)$ d (vsnet-alert 22925).  Our radial velocities, 
from 2015 October and 2018 November, could not disambiguate
candidate frequencies near 13.0 and 14.0 cycle d$^{-1}$. The Vanmunster 
superhump period breaks the ambiguity in favor of 
$1 / P_{\rm orb} = 13.0$ cycle d$^{-1}$.

\subsection{V1024 Per = NSV1436}

\begin{figure}
\includegraphics[height=23 cm,trim = 2.2cm 2cm 1cm 2.8cm,clip=true]{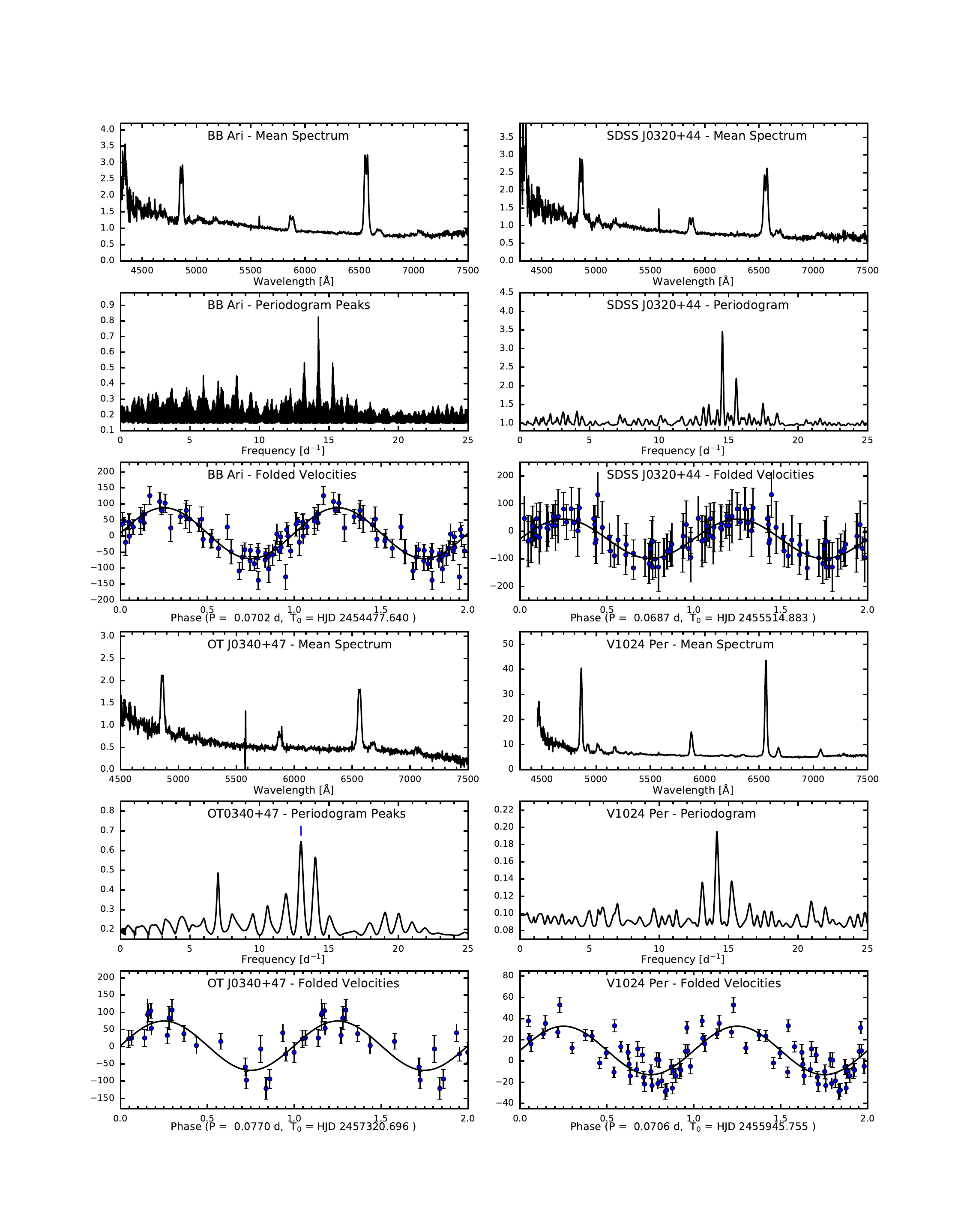}
\caption{Similar to Fig.~\ref{fig:montage1}, but for BB Ari, SDSS J0320+44, 
OT J0340+47, and V1024 Per.  The odd continuum shape for OT J0340+47 is likely
to be an artifact. In the periodogram for that object, the tick mark
in the periodigram shows the frequency selected by the 
superhump period.} 
\label{fig:montage2}
\end{figure}

\citet{kato7} give background information on this object
(Fig.~\ref{fig:montage2}), which was
known as NSV 1436 before it was named V1024 Per. 
They analyzed data from a 2014 September outburst and 
derived superhump periods
$P_1 = 0.072843(14)$ and $P_2 = 0.072403(43)$ d.  
The object has a companion about 3 arcsec to its
south, which is not known to be associated except
along the line of sight.  It lies outside the 
CRTS2 coverage area.

\subsection{V1389 Tau = OT J040659.8+005244}

This object (Fig.~\ref{fig:montage3}) was first reported by \citet{yamaoka08a}.
\citet{kato1} analyzed data from its 2008 superoutburst and
found a nearly constant superhump period of 0.07992(2) day; 
they suggested the source had been caught relatively late
in the outburst, leading to a nearly constant period. 

We have only 24 spectra, 16 from an observing run in 2009 
October and the rest from 2010 November.  The 
combined H$\alpha$ radial velocities 
define $P_{\rm orb} = 0.0780(2)$ d.  The number of cycles 
that elapsed between the runs is unknown, so we estimate
the uncertainty in $P_{\rm orb}$ from fits to the 
individual observing runs.  Our velocities do 
not independently establish the daily
cycle count, but our best period is
slightly shorter than $P_{\rm sh}$, as expected, 
so the alias choice is secure. 

\subsection{Gaia 19emm}

The Gaia Alerts 
Index\footnote{http://gsaweb.ast.cam.ac.uk/alerts/alertsindex} 
notes a brightening of this object 
by $\sim$ 1.5 mag on 2019 Oct.\ 6, and lists it as a 
candidate CV.  The object lies 6.3 arcsec from the Einstein X-ray source 
2E 0431.7+1756, and 13 arcsec from RX J0434.5+1802, and
may be the optical counterpart.

The Gaia light curve shows multiple detections
from 2015 onward, between G = 17 and 18, with one brightening
in 2019 October that triggered the alert, while CRTS DR2 
has it mostly between $V$ = 16.5 and 17.5, with a single 
brightening to $V$ = 15.8 on 2005 Sept. 28.  Spectra obtained
on the McGraw-Hill 1.3 m in 2019 Oct.\ 30 and 31, only 
a few weeks after discovery, showed the broad Balmer and
\ion{He}{1} emission typical of a dwarf nova at minimum
light, at a signal-to-noise inadequate for a radial 
velocity study.  Better and more extensive data obtained
in 2020 January with the 2.4 m and OSMOS yielded an unambiguous radial
velocity period of $123.55 \pm 0.09$ min.  The spectrum
(Fig.~\ref{fig:montage3})
shows the strong, broad emission lines typical of dwarf
novae at minimum light; the FWHM of H$\alpha$ is 1600 
km s$^{-1}$ and its EW is 88 \AA .  The spectrum and 
orbital period are both typical for dwarf nova, but the
outbursts so far recorded are rather weak.

\subsection{V1208 Tau = Tau3}

\citet{motch96} discovered this CV as the optical counterpart
of a ROSAT X-ray source.  It was listed in the \citet{downes} catalog
with the temporarily designation ``Tau3''.  

The CRTS2 light curve shows it typically near 18th magnitude,
but with substantial scatter, going fainter than 19th on 
occasion, and a few points near 21st.  There are at least
nine outbursts, to 14.9 mag near maximum brightness. 
\citet{kato1} gives superhump periods $P_1 = $ 0.070501(32) and
0.070537(27) day for two different outbursts; \citet{kato4}
find $P_2 = 0.070481(66)$ day for a third, and suggest that
all three determinations are actually for Stage C superhumps.
Combining these yields $P_{\rm sh} = 101.5(3)$ min.

The mean spectrum (Fig.~\ref{fig:montage3}) shows single-peaked 
emission lines on a flat continuum.
The radial velocities do not constrain the daily cycle 
count independently, but the well-determined superhump
period selects our best-fitting radial velocity period, 
which is 98.1(2) min.
 


\subsection{ASAS-SN 15pq} 


This was discovered by ASAS-SN on 2015 Sept.~12, and reached
$V \sim 16.3$.  The CRTS DR2 has observations at 
64 epochs that vary from 18.0 to 19.0, with one excursion
to 17.5, so observations so far establish a rather
modest outburst amplitude of $\sim 3$ mag.  We have 
velocities from five observing runs.  The first four
yielded two possible periods  115.5(3) or 108.9(3) min.
The 2020 January data resolved the ambiguity in favor
of the longer period, giving 115.2(3) min for the 
2020 January data alone.  Superoutbursts and superhumps 
have apparently not been reported. 

\begin{figure}
\includegraphics[height=23 cm,trim = 2.2cm 2cm 1cm 2.8cm,clip=true]{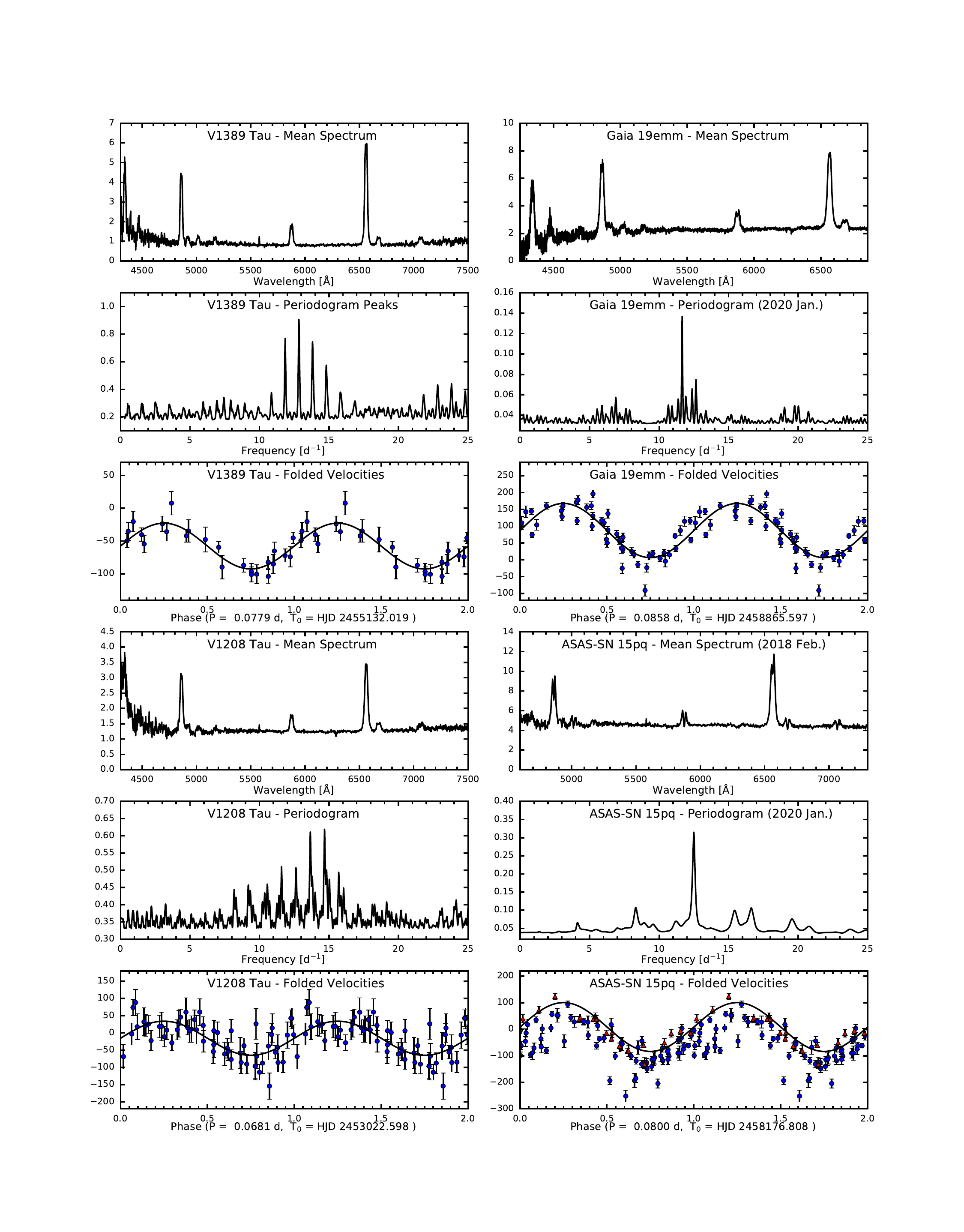}
\caption{Similar to Fig.~\ref{fig:montage1}, but for V1389 Tau, Gaia 19emm, 
V1208 Tau, and ASAS-SN 15 pq.  In the folded velocity plot for ASAS-SN 15pg, the
red triangles are the 2020 January data used to compute the periodogram, and the 
amplitude and zero point of the solid curve are fitted to those points, which 
are systematically high compared to the velocities from the other observing runs.
}
\label{fig:montage3}
\end{figure}

The sinusoidal fit parameters in Table \ref{tab:parameters}
and the periodogram in Fig.~\ref{fig:montage3} are derived
from the 2020 January data.  The phase-folded velocity
plot (also Fig.~\ref{fig:montage3}) also includes points
from the other runs. 

\subsection{Var Cam 06 =  1RXS J055722.9+683219}
\label{subsec:rx0557}

\citet{uemura06} reported the first known outburst of
this object, and 
attribute the discovery to W. Kloehr; they also found 
apparent superhumps.  The object lies 25 arcsec from a 
ROSAT X-ray source (1RXS J055722.9+683219) and is the 
likely optical counterpart.  We will refer to it as
1RXS J0557+685.\footnote{SIMBAD calls this 
``NAME Var Cam 06 = 1RXS J055722.9+683219".}
\citet{uemura10} gave details of this superoutburst
and another some 480 d later; they found $P_{\rm sh}$  
to be relatively short at 
0.05324(2) day, or 76.66 min, indicating that
the orbital period is also short.  The 480-d
interval between superoubursts suggest that they 
are relatively frequent compared
to many short-period SU UMa stars.
\citet{uemura10} reviewed data on
dwarf novae with similarly
short periods, and suggested that 
this object belongs to a natural grouping that
combines very short period, relatively short outburst
recurrence times, and supherhump period excesses that
are larger than expected for the orbital period. 
On this basis they suggested that $P_{\rm orb}$
should be 0.05209(1) d, or roughly 75.01(2) min.

The spectrum (Fig.~\ref{fig:montage4}) is somewhat unusual.  
The \ion{He}{1} lines are atypically strong; 
\ion{He}{1} $\lambda$5876 in particular is a bit
less than half the flux of H$\alpha$.  Although the signal-to-noise
degrades badly toward the blue, \ion{He}{2} $\lambda$4686,
is also clearly detected.  There is no evidence
for a secondary star spectrum, but
the object is relatively faint, with a synthesized 
$V$ magnitude near 19.3, limiting the signal-to-noise
ratio to $\sim 20$ per 2 \AA\ pixel in the best-exposed
parts of the spectrum.  
The spectrum is reminiscent of, but less extreme than, 
that of the dwarf nova SBS 1108+574 \citep{carter13}, 
which has $P_{\rm orb} = 55.3$ min, and a 
ratio of \ion{He}{1} $\lambda 5876$ to H$\alpha$ of 
0.81.  In addition, both that object and the 
present one show weak emission centered on 6356 \AA ,
which \citet{carter13} attribute to a blend
of \ion{Si}{2} $\lambda\lambda$ 6347 and 6371.  SBS 1108+574
shows broad absorption wings in the higher Balmer
lines that are attributed to the white dwarf in the system, 
but these are not evident here.

The measured orbital period, 75.34(38) min, agrees
within the mutual uncertainty with the period predicted
by \citet{uemura10}.

\subsection{SDSSJ075117.00+100016.2}

\citet{wils10} identified this object (hereafter SDSS J0751+10) 
as a CV.  Spectra 
from five observing runs in 2010 and 2011 did not reveal
an unambiguous period, but a more concerted effort in 
2020 Jan.~showed the period to be 85.28(7) min, which 
also fits the earlier velocities acceptably well.  As with 
ASAS-SN 15pq, the parameters
in Table \ref{tab:parameters} and the periodogram (Fig.~\ref{fig:montage4})  
are from the 2020 January data alone, but the folded velocity plot
includes all the data.


\subsection{SBSS 0755+600} 

\begin{figure}
\includegraphics[height=10 cm,clip=true]{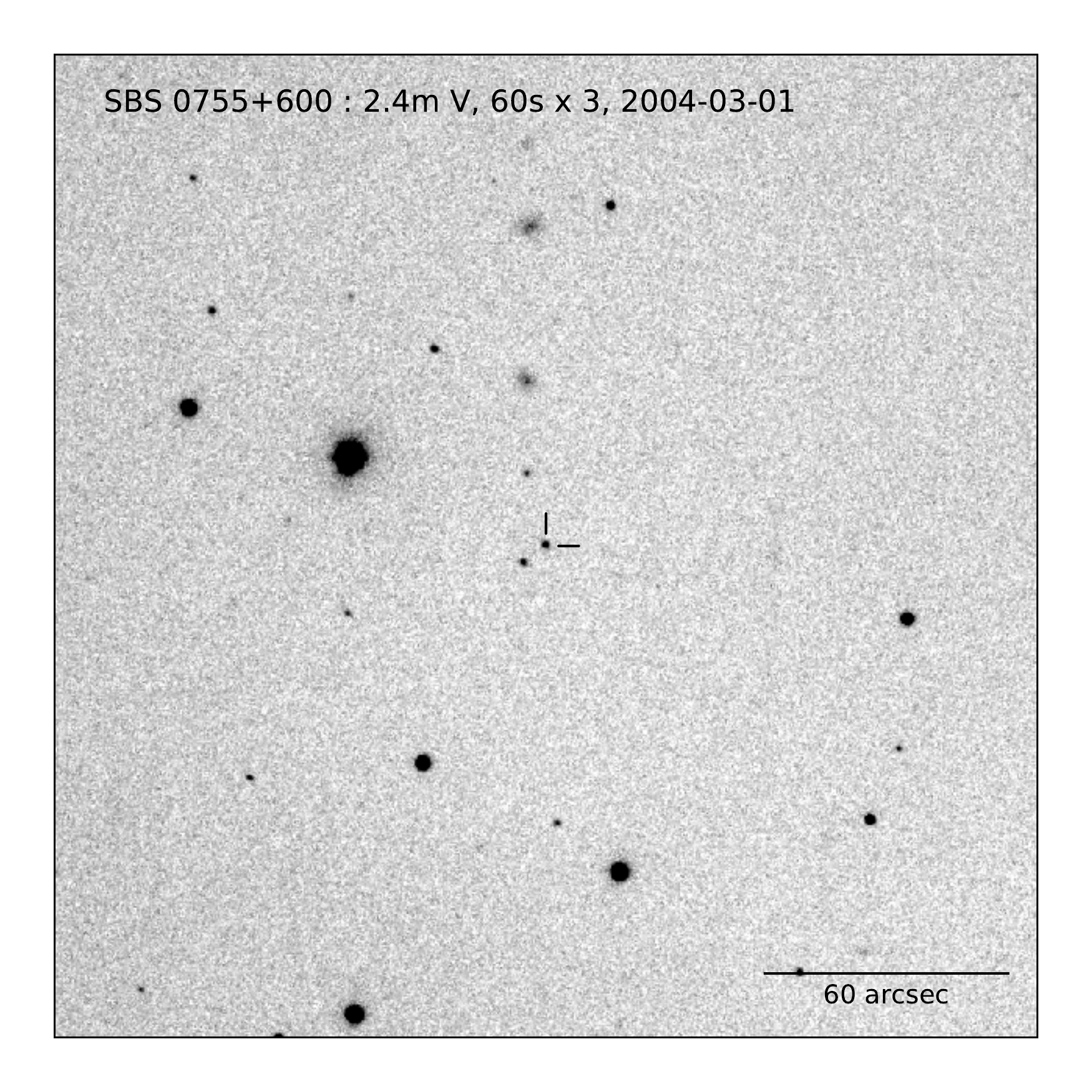}
\vspace{0.4 truein}
\caption{Finding chart for SBSS 0755+60, from the median of three 60 s $V$-band 
exposures taken with the MDM 2.4m telescope on 2004 March 1 UT.  North 
is at the top and east is to the left, and the scale is indicated.}
\label{fig:findingchart}
\end{figure}

\begin{figure}
\includegraphics[height=23 cm,trim = 2.2cm 2cm 1cm 2.8cm,clip=true]{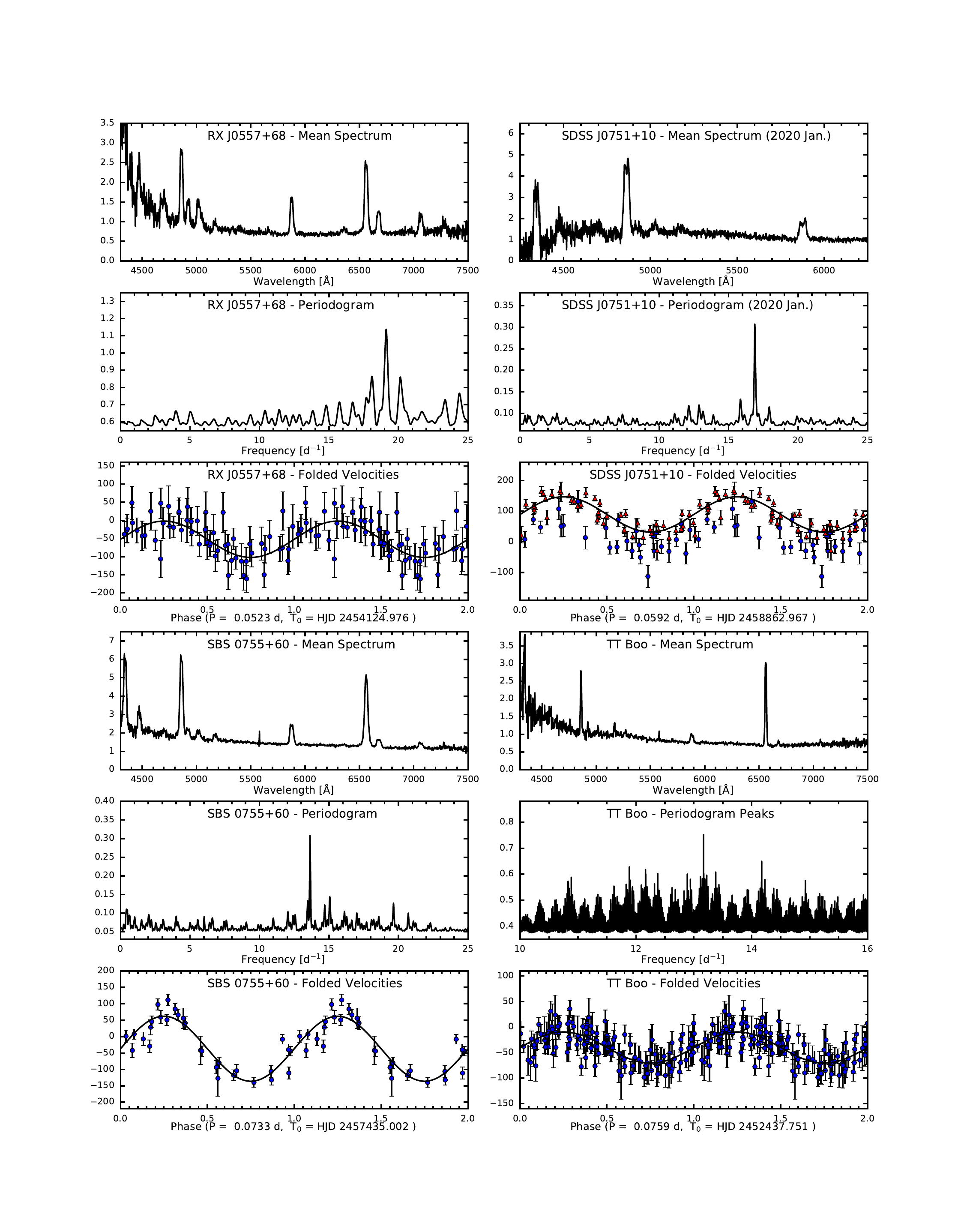}
\caption{Similar to Fig.~\ref{fig:montage1}, but for 
1 RXS J05573+865, 
SDSS J0751+10,
SBSS 0755+600, 
and 
TT Boo.
The downturn at the blue end of the spectrum of SDSS J0751+10
is unlikely to be real.  For the same star, the 2020 January
velocities are shown with red triangles, and again appear to be
systematically high compared to the velocities from other runs.
A 3-point boxcar smoothing has been applied to the (oversampled) spectrum.}
\label{fig:montage4}
\end{figure}

\citet{stepanian99} took spectra of a sample of 
objects from the Second Byurkan Survey and 
classified this as a CV.  
It appears in outburst on the POSS-I images (obtained
through the Digital Sky Survey at STScI),
which were taken 1954 January 5.
The outburst images show the CV 
slightly brighter than a field star near $\alpha = 7^{\rm h} 59^{\rm m}
24^{\rm s}.00, \delta = +59^{\circ} 52' 31''.0$, about
80 arcsec south and somewhat west of the CV; 
the APASS catalog lists $V = 15.7$ for this
star.  The finding chart in the \citet{downes} catalog is incorrect --
the CV is slightly northwest of the object they mark. 
Fig.~\ref{fig:findingchart} shows the correct star.  
The CRTS light curve has 120 points, but no outbursts; 
the magnitude reaches 17.39 and sometimes fades to 
fainter than 19th, reaching 20.51 on a single occasion. 
(See Fig.~\ref{fig:montage4}.)

\subsection{TT Boo}

This is among the longer-known objects considered here
\citep{chernova51}.  It is 
an SU UMa star with numerous well-observed
superoutbursts \citep{kato1, kato3, kato4} giving
$P_{\rm sh 1}$ near 0.0780 d, or 12.82 c d$^{-1}$.
The CRTS2 has 228 points covering
about 9 years, and shows the star mostly near 19th magnitude,
with three outbursts, one of which reaches 13.34. 

We have 127 velocities taken on many runs over a span 
of 5552 days.  The daily cycle count is not definitive, 
but strongly favors a value near 13.168 c d$^{-1}$, as 
expected from the superhump period.  We nearly have a
nearly unambiguous run-to-run cycle count, the only remaining 
ambiguity being a splitting at one cycle per 5000 days.  
The small period uncertainty quoted in Table~\ref{tab:parameters} 
reflects this. (See Fig.~\ref{fig:montage4}.)


\subsection{QZ Lib = ASAS J153616-0839.1}

This object was discovered in outburst in 2004 by ASAS.
It showed superhumps at $P_1 = 0.064602(24)$ day \citep{kato1}.
Apparently no outbursts have been observed since then, 
suggesting a WZ Sge classification.  Our mean spectrum 
(Fig.~\ref{fig:montage5}) shows
strong, relatively narrow Balmer emission lines and  
apparent broad absorption wings at H$\beta$, apparently
from an underlying white dwarf contribution.

Our radial velocities do not span enough hour angle range to 
independently determine a period.  However, only one of our
daily cycle-count aliases -- the second-best one -- 
is consistent with the well-determined $P_{\rm sh}$,
so we adopt this, and find $P_{\rm orb} = 0.06411(7)$ d.
Taking $P_1$ from \citet{kato1} as 
$P_{\rm sh}$ gives $\epsilon = 0.0071 \pm 0.0013$, suggesting
a very low mass ratio.  

After our data were collected,
\citet{pala18} published a detailed study of QZ Lib.
They independently determined
$P_{\rm orb} = 0.06436(20)$, consistent with our determination but
with a slightly larger formal uncertainty.
They calculate a mass ratio $q = M_2/M_{\rm WD} = 0.040(9)$,
find $T_{\rm WD} = 10 500 \pm 1500 $K, 
and conclude that QZ Lib is likely to have passed through 
its period minimum, making it a period ``bouncer''.

\subsection{CSS170517:155156+1453334} 



This object (hereafter CSS 1551+14)
was discovered at magnitude 15.64 on 2017 May 17.  
The CRTS DR2 has 302 epochs over a span of 3000 days,
and shows variation between 17.2 and
19.0 at minimum light, and two outbursts to around 15.5, 
that escaped notice at the time.  

Our velocities (Fig.~\ref{fig:montage5}) are from 2017 June, not long after the discovery.
The observations are a bit sparse; the velocities strongly favor a period
of 100.3(3) min, but aliases are not entirely excluded.


\subsection{SDSS J155720.75+180720.2} 

\citet{szkodyvii} discovered this object (hereafter SDSS 1557+18) 
in the SDSS; it lies 8 arcseconds
from the ROSAT X-ray source 1RXS J155720.3+180715 and is
the likely optical counterpart.
\citet{szkodyvii} derived a preliminary period near 2 hours from a short
run of radial velocity data.  In its CRTS2 light curve it appears
consistently near 18th magnitude, but at 14.8 mag in 
2007 June 13 and 16.2 on 2008 June 9.  A further outburst
was reported by Jeremy Shears on 2015 July 30 (vsnet-alert 18914),
and another was found by CRTS in 2016 March (vsnet-alert 19602);
during this last outburst J. Hambsch established $P_{\rm sh}$ = 
0.08538(6) d.  (See Fig.~\ref{fig:montage5}.)


\subsection{IX Dra}
\label{subsec:ixdra}

This dwarf nova outbursts frequently; \citet{olech04} give
3.1(1) and 54(1) d for the mean cycle and supercycle lengths,
respectively.  \citet{olech04} obtained time-series photometry
during a 2003 September superoutbust and established $P_{\rm sh} = 
0.066968(17)$ day, or 96.43(2) min.  Intriguingly, their
photometry showed a second periodicity at 0.06646(6) day, or
95.70(8) min, only
0.76 per cent shorter than the superhump.  They pointed out that
identifying this second period with $P_{\rm orb}$ implies a 
very small mass ratio.  If confirmed, this would make the
system an excellent candidate for a post-bounce CV (see the 
Introduction). 

\begin{figure}
\includegraphics[height=23 cm,trim = 2.2cm 2cm 1cm 2.8cm,clip=true]{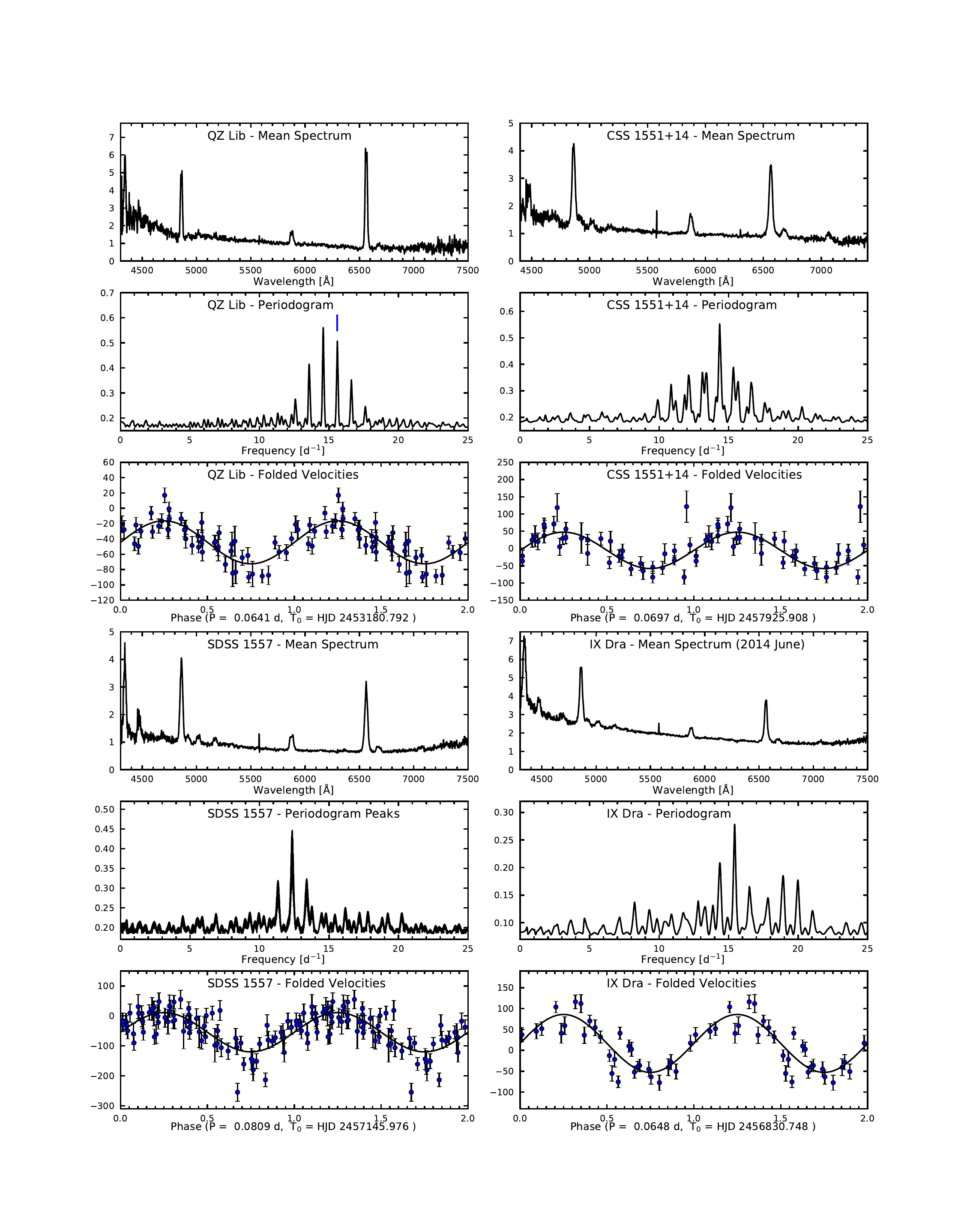}
\caption{Similar to Fig.~\ref{fig:montage1}, but for 
QZ Lib, 
CSS 1551,
SDSSS 1757, 
and 
IX Dra.   In the periodogram for QZ Lib, 
the tick mark indicates the  adopted orbital 
period (based on $P_{\rm sh}$).
In SDSS 1557+18, the upturn in the continuum 
longward of H$\alpha$ is probably an artifact.
}
\label{fig:montage5}
\end{figure}

The very short outburst cycle made it difficult to obtain
a usable velocity time series; on multiple
visits over the years, we found it either in full outburst 
with absorption lines, or well above minimum light with relatively 
weak emission.  In 2014 June we finally 
found the star in a fairly low state with 
tractable emission lines (Fig.~\ref{fig:montage5}).  
We obtained 29 radial velocities on three
successive nights, after which the source returned
to outburst.  From the velocities we find $P_{\rm orb} = 
0.06480(16)$ d, or 93.31(23) min, 3.3 per cent shorter
than $P_{\rm sh}$.  This is consistent with the 
$\epsilon$ - $P_{\rm orb}$ relation, so the system 
is (perhaps disappointingly) apparently not post-bounce.

\subsection{ASAS-SN 13at = MASTER OT J182122.59+614854}

This object was discovered independently by the ASAS-SN and
MASTER surveys \citep{atel5168,atel5182}.  Our observations
are from 2013 Sept., 2014 June, and 2018
June, and define $P_{\rm orb} = 0.0790(2)$ d without
ambiguity in the daily cycle count, although the number 
of cycles elapsed between observing runs is unknown.  
(See Fig.~\ref{fig:montage6})


\subsection{KIC 11390659}

\citet{howell2013} obtained spectra of this {\it Kepler} object
{\footnote{This object is listed in SIMBAD as 2MASS J18583091+4914326.}
showing strong double-peaked emission lines typical of a dwarf nova at 
minimum light.  A short radial-velocity time series indicated a period
of 106-109 min.  We observed the source on three successive nights
in 2018 May with the McGraw-Hill 1.3m telescope, and found
an unambiguous period of 110.80 $\pm 0.17$ min, apparently consistent
with that found by \citet{howell2013} (see Fig.~\ref{fig:montage6}).
The Kepler light curve showed only irregular variations, but
the ASAS-SN survey found an outburst on 2013 July 3.
Photometry during that outburst did not reveal any superhumps
(Pavol A. Dubovsky, Tamas Tordai and Tonny Vanmunster, vsnet-alert  
18729).  A subsequent outburst was observed in 2017 May by 
Tadashi Kojima (vsnet-alert 20982), again without any superhump
detection.  


\subsection{RX J1946.2-0444}

This object was discovered by \citet{motch98} in the Rosat Galactic Plane Survey,
and catalogued by \citet{downes}.  Its position 
is not covered in CRTS DR2, and apparently nothing about its variability
has appeared in the literature.  

We observed this on three observing runs in the 2007.
Although outbursts have not been observed, 
the mean spectrum (Fig.~\ref{fig:montage6}) appears entirely typical 
of dwarf novae at minimum light.  Our observations were sufficient
to establish an unambiguous cycle count for the data set, 
allowing us to establish a relatively precise $P_{\rm orb} = 107.46(1)$ min.

\begin{figure}
\includegraphics[height=23 cm,trim = 2.2cm 2cm 1cm 2.8cm,clip=true]{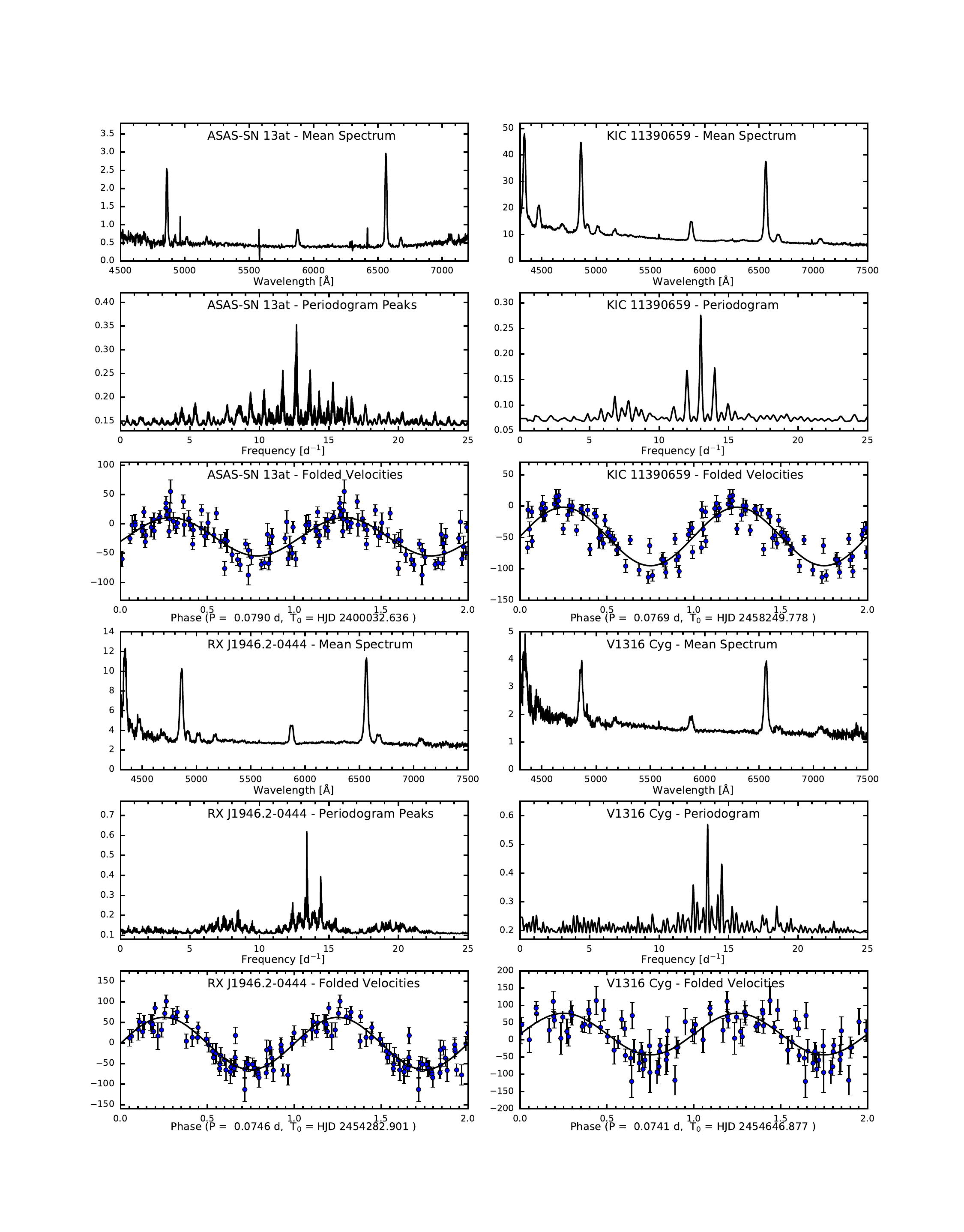}
\caption{Similar to Fig.~\ref{fig:montage1}, but for 
ASAS-SN 13at, 
KIC 11390659, 
RX J1946.7-0444, and
V1316 Cyg. 
The velocities do not determine the period of ASAS-SN 14cl 
unambiguously; the adopted period, chosen for
consistency with the superhump period, is indicated in the periodogram.
}
\label{fig:montage6}
\end{figure}


\subsection{V1316 Cyg}

\citet{romano67} discovered this variable star (first designated GR 141)
and later classified it as a dwarf nova \citep{romano69}.  \citet{boyd08}
detected superhumps at 0.07685(3) d, which transitioned to 0.07654(2) d
later in the outburst. Fig.~\ref{fig:montage6} summarizes our data.

\subsection{ASAS-SN 14ds}

ASAS-SN detected an outburst 
from this source 2014 July 08 \citep{asn14ds}, designated
it ASAS-SN 14ds, and identified it with the
X-ray source 1RXS J204455.9$-$115151.  
Denisenko (vsnet-alert 17478) drew attention to at least
one previous outburst, and noted that the source varies
by about 1 mag in quiescence.  However, \citet{hardy17} obtained
a $\sim 5$ hr light curve during outburst and found
no eclipse.  We obtained an exploratory
spectrum with the Hiltner telescope and OSMOS in 2016
August, which appeared typical of a dwarf nova, and much
more extensive modspec data 2018 August and September with
the McGraw-Hill 1.3m and Hiltner 2.4m respectively that
unambiguously indicated a period near 2.53 hr, squarely
in the period gap (Fig.~\ref{fig:montage7}).  The weighted average of the periods
from separate fits to the August and September data
is 0.1058(3) d.  The two runs are close enough together
that there are relatively few viable choices of cycle count over
the inter-run gap; the resulting periods can summarized
as 
\begin{equation}
P_{\rm orb} = {25.553 \pm 0.007\ {\rm d} \over N},
\end{equation}
where $N = 242 \pm 2$ is an integer.


\subsection{V444 Peg = OT J213701.8+071456}

This variable star was discovered on 2008 Nov.~6 by \citet{yamaoka08b}.  The 
441 points in its CRTS2 light curve show it mostly fluctuating from 
18th to 19th magnitude, with an outburst to 12.8 mag in 2012 September.
there are also three points taken 2008 Nov.~21 near 16.6 mag,
apparently near the end of the discovery outburst.  
\citet{kato1} gives $P_1 = 0.099451$ for the 2008 outburst
(with no error estimated), and $P_1 = 0.097645(52)$ for the 
2012 outburst.  

The spectrum (Fig.~\ref{fig:montage7}) is unusual in that it shows a contribution
from an M-dwarf secondary star; the signal-to-noise ratio does
not permit an accurate classification of the secondary, but we
estimate it to be M3 $\pm$ 1.5 subclasses from comparison with 
library spectra.

We find $P_{\rm orb}$ = 2.222(4) hr, at the lower edge of
the period gap; the superhump period excess $\epsilon$ is 
then 0.073 using the value from the 2008 outburst,
and 0.054 for the shorter $P_{\rm sh}$ in the 2012 outburst.



\subsection{ASAS-SN 14cl} 

\citet{atel6233} announced the discovery of this transient,
which reached $V = 10.66$.
\citet{kato7} give details of the outburst, and
find superhump periods $P_1 = 0.060008(13)$ and $P_2 = 0.059738(14)$ 
d.  The object was caught early enough that so-called `early superhumps' 
-- a photometric modulation that is apparently at the orbital period --
were detected, yielding an estimated $P_{\rm orb} = 0.05838$ d.  

Our spectrum (Fig.~\ref{fig:montage7}) shows a relatively steep blue continuum and deep, broad
$H\beta$ absorption flanking the emission line, both of which 
indicate a strong contribution from the white dwarf photosphere.
The $H\alpha$ emission line had a low velocity amplitude, and we 
cannot independently determine an unambiguous $P_{\rm orb}$.  However,
one or our candidate periods, 0.05859(4) d, is apparently consistent
with the early-superhump period, for which no uncertainty is quoted.  


\begin{figure}
\includegraphics[height=23 cm,trim = 2.2cm 2cm 1cm 2.8cm,clip=true]{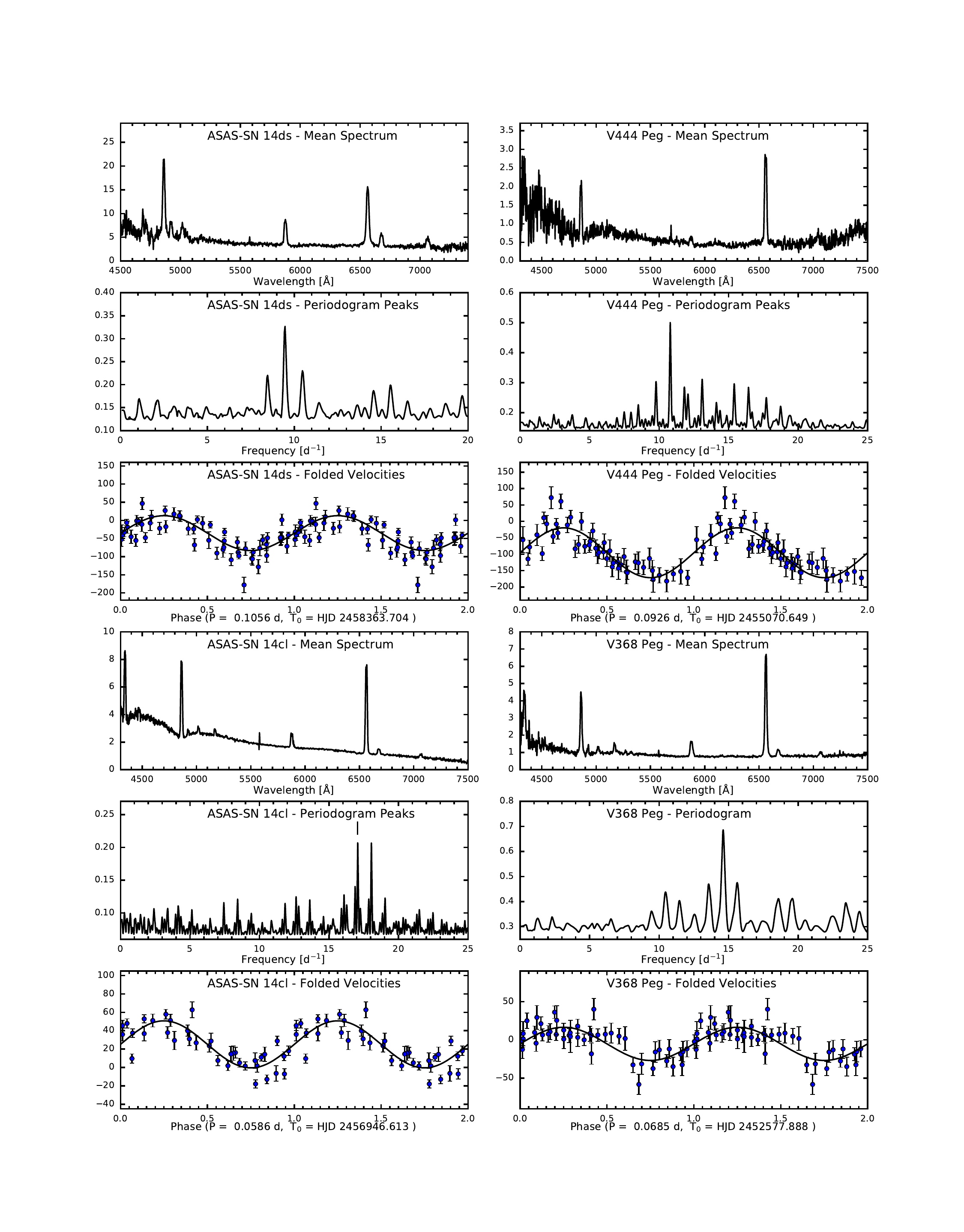}
\caption{Similar to Fig.~\ref{fig:montage1}, but for 
ASAS-SN 14ds, 
V444 Peg, 
ASAS-SN 14cl, 
and 
V368 Peg.
}
\label{fig:montage7}
\end{figure}


\subsection{V368 Peg = Antipin V63}

\citet{antipin99} discovered this star on archival plates, which 
showed 7 outbursts reaching as bright as 13.3 mag.  
\citet{kato1} and \cite{kato2} describe the superhump behavior
in several different outbursts and found superhump periods
near 0.07039 d.  We find the amplitude of the H$\alpha$ emission 
velocity modulation to be quite low ($K = 22 \pm 5$ km s$^{-1}$),
but were still able to determine $P_{\rm orb} = 0.0685(3)$ d from
data taken mostly on two successive nights in 2002 October 
(Fig.~\ref{fig:montage7}).


\subsection{IPHAS 2305} 

This source was originally discovered as an H$\alpha$ 
emission source by the INT/WFC Photometric H$\alpha$
Survey (IPHAS), described by \citet{witham08}. 
{It is a possible optical counterpart of the ROSAT
X-ray source 1RXS J230538.2+652155.}
A superoutburst in 2015 was analyzed by
\citet{kato8}, who found $P_{\rm sh} = 0.0727$ 
d.  We cannot independently find an unambiguous
period from our radial velocities, which date from 
three nights in 2008 September, but the period
that fits our data best -- 0.0702(3) d -- is 
nicely consistent with the superhump period, and we
are confident that it represents $P_{\rm orb}$.
It is marked in the periodogram (Fig.~\ref{fig:montage8})


\subsection{NSV14652 = CRTS J233848.7+281955}

\citet{kato1} found $P_{\rm sh} = 0.081513(16)$ d 
for a 2004 outburst of this SU UMa star, which SIMBAD 
lists by its CRTS designation given above.  The 
Catalina Sky Survey detected an outburst on 2009 June 27.
The CRTS DR2 light curve shows it mostly around 18th magnitude,
with only a few outbursts, the brightest reaching 15th.
\citet{szkodyfollowup} published a spectrum of the source
in decline, showing Balmer emission lines. Our spectra
(Fig.~\ref{fig:montage8}) appear similar.

\begin{figure}
\includegraphics[height=23 cm,trim = 2.2cm 2cm 1cm 2.8cm,clip=true]{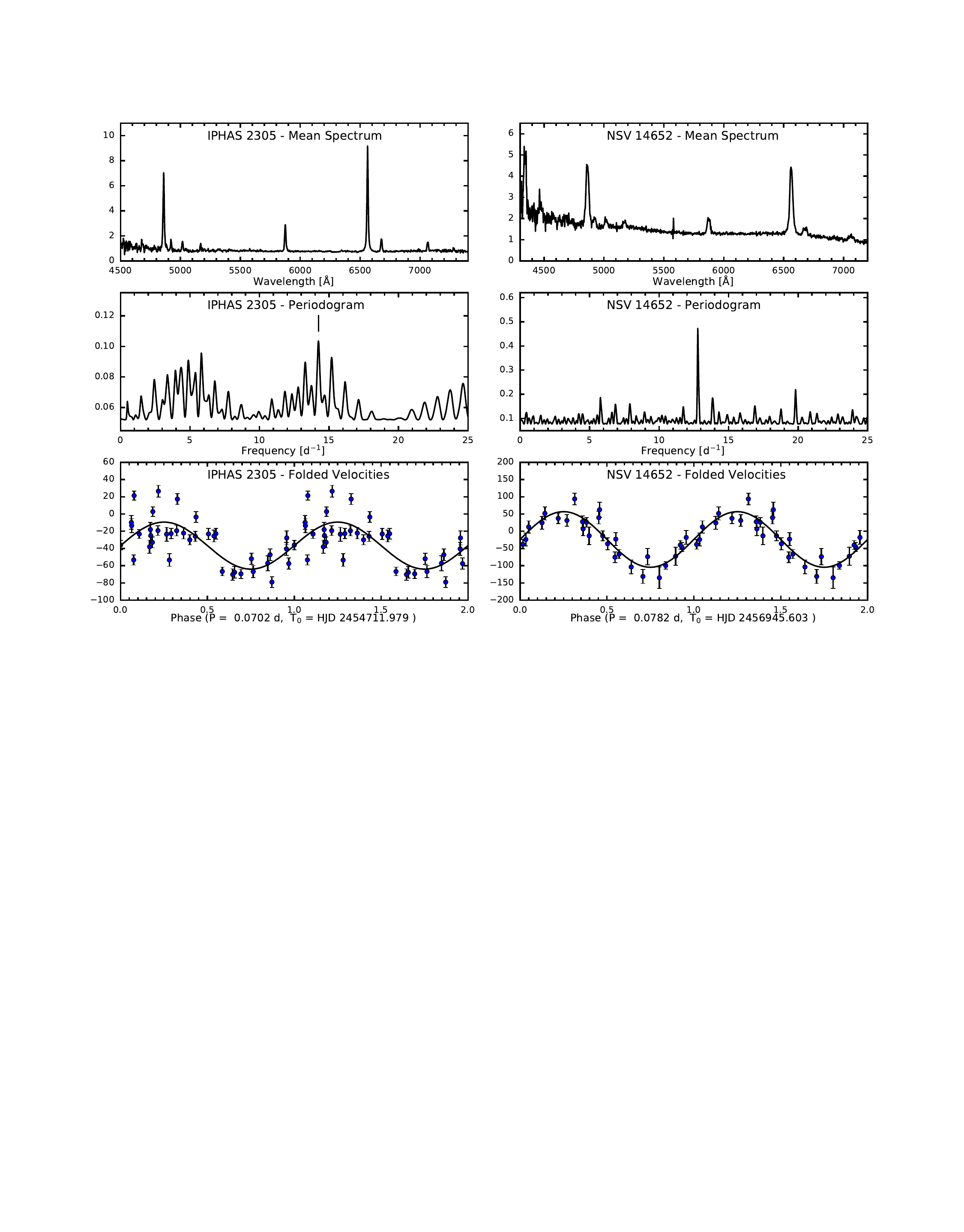}
\caption{Similar to Fig.~\ref{fig:montage1}, but for IPHAS 2305
and NSV14652.  The period adopted for IPHAS 2305, based on the 
superhump period, is indicated in the periodogram.
}
\label{fig:montage8}
\end{figure}

\clearpage

\section{Discussion}

Dwarf novae in this period range often show superoutbursts 
and superhumps.  We measure $P_{\rm orb}$ independently 
of $P_{\rm sh}$ here, so we can compute empirical values 
of the superhump period excess $\epsilon$ for those stars 
that also have $P_{\rm sh}$.

\subsection{The $\epsilon$-$P_{\rm orb}$ relation.}

The superhump period excess $\epsilon = 
(P_{\rm sh} - P_{\rm orb}) / P_{\rm orb}$ should be
a good proxy for the mass ratio $q = M_2 / M_{\rm WD}$; 
\citet{patt05} calibrates this relationship empirically 
and approximates it as
\begin{equation}
\epsilon = 0.18 q + 0.29 q^2,
\end{equation}
which we adopt here. 

Short-period CVs are
thought to evolve to shorter periods until they reach a period 
minimum, after which they should evolve toward longer periods.
On the $\epsilon$-$P_{\rm orb}$ diagram, one expects the
`post-bounce' systems to be objects with $P_{\rm orb}$ somewhat 
longer than the minimum period, and with $\epsilon$ 
(and hence $q$) 
smaller than expected, indicating a very low-mass
secondary.  
\citet{pattmurmurs} lists 22 candidate post-bouncers,
based on a variety of criteria.  Since then, more
have been proposed (e.g. \citealt{kimura18, nakata14, mcallister17,
pala18}).

In a remarkable series of papers \citep{kato1, kato2, kato3, kato4, kato5,
kato6, kato7, kato8, kato9, kato10}, T. Kato and collaborators
present extraordinarily extensive time series 
photometry of outbursting SU UMa stars.  The papers also 
tabulate orbital periods, where known.  The shorter-period
objects in the sample, which tend to be seldom-outbursting
WZ Sge stars, often show so-called early superhumps, which
appear early in the outburst before the development of
normal superhumps.  These appear to be reliable indicators
of the orbital period itself.  Therefore, many of the 
shorter-period objects included in these papers have reasonably
reliable $P_{\rm orb}$ from photometry alone; others have 
spectroscopic periods.  

Fig.~\ref{fig:qvsp} shows the mass ratios $q$ inferred from 
the superhump period excesses $\epsilon$ (using, specifically,
the average `Stage A' superhump period, or $P_1$ in 
their notation) plotted against $P_{\rm orb}$.   The curve 
shown is a polynomial for $P_{\rm orb}$ as a function of 
$q$, constructed by fitting the points from the 
Kato papers, rejecting points that missed
the curve by more than 0.005 d ($\sim 7$ min), and 
iterating twice, in order to represent the apparent
`ridgeline' of the distribution.  Its analytic
form is 
\begin{equation}
P_{\rm orb} = 0.07276688 - 0.49012845\, q + 3.89878883\, q^2 - 5.65565187\, q^3,
\end{equation}
where $P_{\rm orb}$ is in days, and the range of validity is
as shown in Fig.~\ref{fig:qvsp}.
The curve reaches a minimum $P_{\rm orb}$ near 80 min,
with $q \sim 0.075$, but individual systems reach 
significantly shorter periods.  
Fig.~\ref{fig:qvsp_detail} is a magnified version
of the turnaround region.  Objects in the shaded
region are evidently strong period-bounce candidates;
Table \ref{tab:bouncers} gives more detail
on these objects.  

\begin{figure}
\includegraphics[height=12 cm,clip=true]{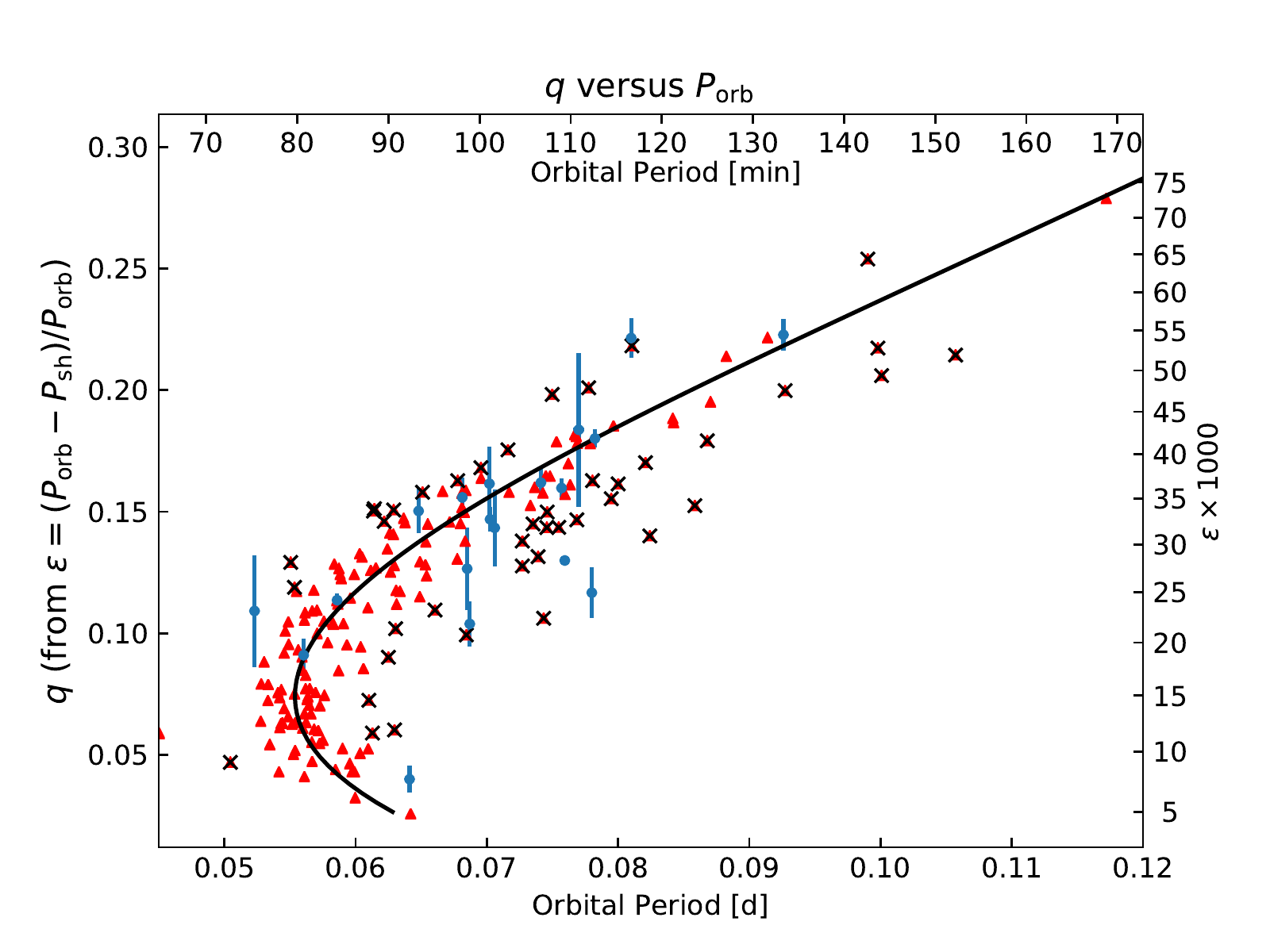}
\vspace{0.4 truein}
\caption{Mass ratios $q$ inferred from superhump period
excesses, plotted against $P_{\rm orb}$.  Small red
triangles are from the series of papers by Kato et al.; 
blue dots with error bars are from the present work.  The black curve
is an iterative fit to the red points (see text).
Points that were excluded by the iterative fit are 
marked with a black x-cross.  The scale on the right
axis gives the superhump period excesses corresponding
to $q$. 
}
\label{fig:qvsp}
\end{figure}

\begin{figure}
\includegraphics[height=12 cm,clip=true]{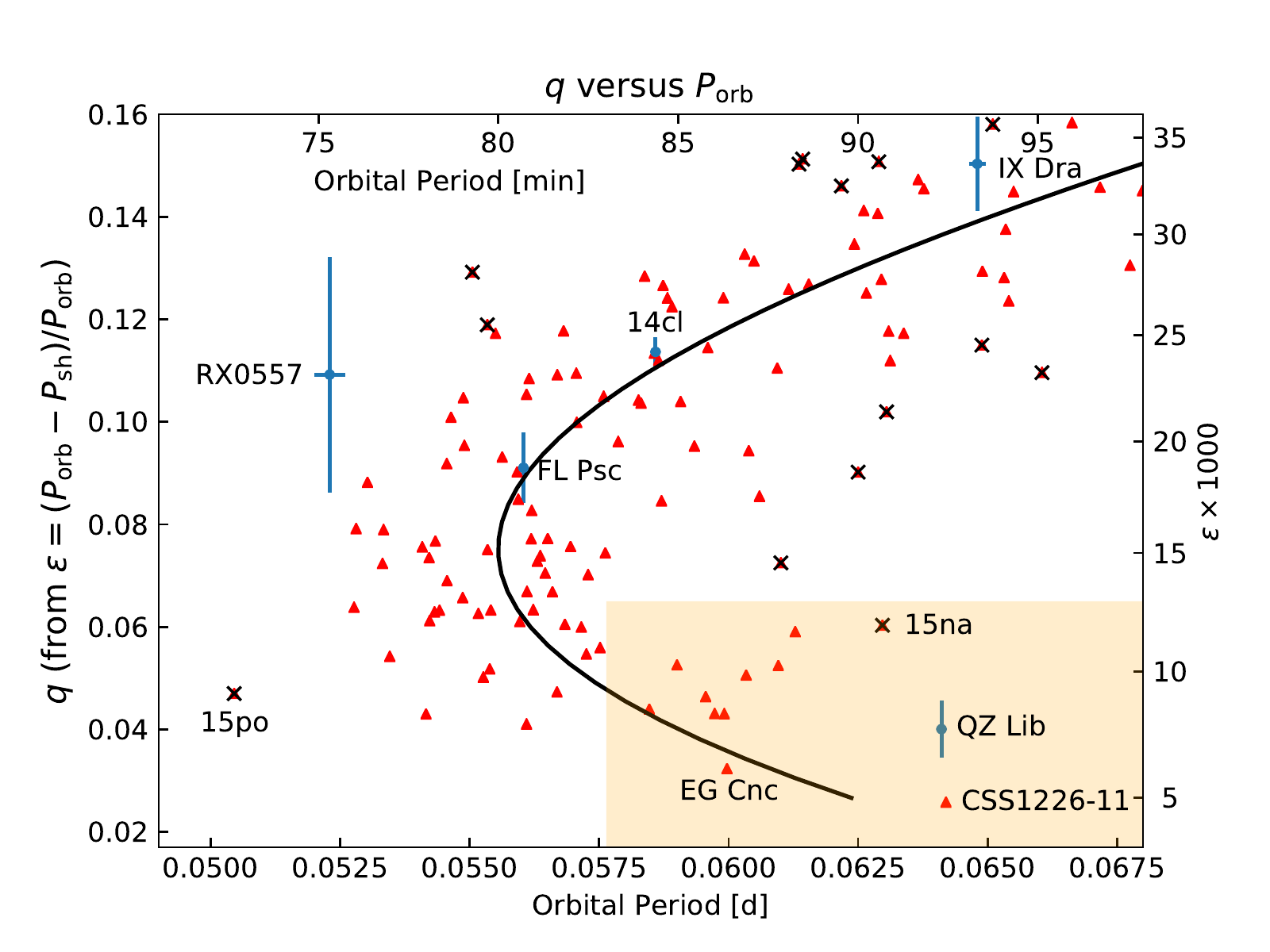}
\vspace{0.4 truein}
\caption{Magnified view of the turnaround portion of 
Fig.~\ref{fig:qvsp}, with selected stars labeled.
Names are abbreviated where necessary; 14cl,
15po, and 15na are from the ASAS-SN survey.
The shaded portion shows strong bounce candidates.}
\label{fig:qvsp_detail}
\end{figure}

The new orbital periods presented here contribute
relatively little to this analysis.  The most notable
are QZ Lib, for which the uncertainties are reduced,
and RX J0557+68, which has a short period
and yet a fairly large $q$, though with substantial
uncertainty.  \citet{uemura10} noted the unusual nature
of this object already (see Section \ref{subsec:rx0557}).
The newly-determined period of IX Dra is nicely consistent with the
upper (pre-bounce) part of the evolutionary trajectory
(Section \ref{subsec:ixdra}).

\subsection{Do all `dwarf novae' undergo outbursts?}
\label{subsec: lurkers} 

As noted in the Introduction, there are several discovery
channels for CVs.  Most CVs were historically discovered
as dwarf novae in outburst, and this has remained the case,
especially as the cadence and depth of all-sky surveys
has dramatically improved.  However, one object discussed
here, RX J1946.2$-$0444, is not known to outburst, yet
appears otherwise similar to the rest of the sample.  Since
short-period CVs tend to have low luminosities, they
are relatively easy to overlook, which leads to the 
the question as to how many more non-outbursting objects 
remain undiscovered.

One way to attack this question compares the rate
of new discoveries {\it versus} re-discoveries in a 
survey such as CRTS.  From an analysis of discovery 
rate versus time, \citet{breedt14} suggest that the
Catalina surveys had found most of the high-accretion-rate
dwarf novae within their footprint and magnitude limits,
but point out that low-accretion rate systems would
still remain undiscovered.  ASAS-SN is still discovering 
new dwarf novae, in part because it covers low Galactic 
latitudes that CRTS did not. 

It is possible that these surveys could
still be missing a population of systems that physically
resemble dwarf novae at minimum light, but undergo 
outbursts on extremely long time scales (or possibly never).
There is no obvious reason why such objects -- 
which we will call {\it lurkers} -- 
could not exist.  Lurkers would necessarily 
have low mass transfer rates $\dot{M}$; if low enough, this
could stretch the outburst interval to be arbitrarily
long.  Alternatively, some residual form of disk
viscosity might be sufficient to maintain a steady
state, again at sufficiently low $\dot{M}$.


To search for lurkers, we examined 
the sample of 284 CVs from SDSS, published
in a series of annual papers by Szkody and others
\citep{szkodyi,szkodyii,szkodyiii,szkodyiv,szkodyv,
szkodyvi,szkodyvii,szkodyviii}.  Importantly, 
these were {\it not} selected because of their variability, but 
rather by color and confirmed with follow-up spectroscopy,

To begin, we examined 
the SDSS spectra and selected those that appeared
consistent with dwarf novae at minimum light.
The classification criteria were (a) emission lines with 
relatively high equivalent width; 
(b) continua that did not slope up too steeply into
the blue, unless white-dwarf absorption wings were
visible around H$\beta$; (c) relatively broad 
emission lines; and (d) weak or absent \ion{He}{2}
emission. These criteria largely eliminate
novalike variables, so-called `polars' or AM Herculis
stars, and systems for which the CV classification
is questionable.  This left 179 of the original 284 objects.

We divided our candidate dwarf nova spectra into
three non-overlapping types.  Objects with apparent
late-type secondary contributions -- which 
usually have orbital periods above $\sim 5$ hr --
were classified as DN-2.  Objects showing a white
dwarf contribution -- which must have 
low disk luminosities, and often have orbital periods 
below 3 hours -- were designated DN-W.   All others 
were simply DN.  

Thirty-seven of the sample were rediscoveries of 
known outbursting dwarf novae.   Eliminating these 
left 142 objects.

Our task was to find how many of these 142 have 
been found to outburst.  This called for a 
complicated search, because individual outbursts
are seldom written up as papers. We proceeded
as follows.

As noted earlier, the archived vsnet-alert emails
are a rich source of information on these stars,
especially announcements of outbursts.  To make
these more accessible we scanned the email subject
lines with a regular expression-based program that
found variable star names, coordinate-based names, 
and names derived from, e.g., the ASAS-SN survey, and
used the results to match the subject lines to a master 
list of CVs.  We then checked the 142 remaining 
candidates for vsnet-alert messages, and found
outburst notices for 81 of them, leaving 61
candidates remaining.  The texts of the
vsnet messages most often credited the Catalina
surveys for the outburst trigger. 

We next downloaded the CRTS-DR2 light curves for the
remaining 61 candidates, and examined them for 
outbursts of greater than $\sim 1$ magnitude.
This test was intentionally not stringent, because
we were trying to isolate true non-outbursters.
This yielded outbursts, or candidate outbursts, 
for another 26 objects from the sample.
Four more were found outbursting in the Palomar
Transient factory database, hosted at the 
NASA/IPAC Infrared Science Archive.  We then turned
to the DASCH interface (Digital Access to a 
Sky Century @ Harvard)\footnote{https://projects.iq.harvard.edu/dasch/about}
which revealed outbursts of three more.  We 
examined the entries in the American
Association of Variable Star Observers (AAVSO)
VSX database of variable stars, which eliminated
one, and finally examined light curves for the 
locations from the ASAS-SN archive hosted at
Ohio State, eliminating a single object.
Finally, we examined the PAN-STARRS light 
curves and found no outbursts, which is not surprising 
given the relatively sparse time sampling.  

After this, 22 objects remained.  They are listed in
Table \ref{tab:lurkers}, and broken down by type
in Table \ref{tab:lurkertype}.  
We draw the following conclusions.

First, the number of non-outbursting objects is
remarkably small, and not compatible with a large
population of lurkers.  While some of the events
we called `outbursts' may have been questionable, it is 
also possible that some outbursts were missed.  If 
we take the fraction of lurker candidates to known 
outbursters -- 22/179, or about 12 per cent -- to 
be representative of the CV population as a whole, 
it is likely that only 10 or 20 per cent of the 
DN-like objects within a few hundred parsecs outburst
so infrequently as to remain 
undiscovered.  \citet{pala19} reach broadly similar 
conclusions using different methods.

It should be noted that this limit applies 
only to objects that appear similar to quiescent
dwarf novae, that is, that have $\dot M$ high enough
to generate the broad emission lines that are the
signature of a DN near minimum.  Mass transfer might
turn off for an extended period, as in the disrupted 
magnetic braking scenario invoked to explain the 
period gap.  
Also, the SDSS CV sample was selected
by their blue colors; very old and slow-accreting
systems might have white dwarfs that were too 
red to be included.   

Table \ref{tab:lurkers} also demonstrates a strong
over-representation of systems with significant
white-dwarf contributions in their spectra 
(DN-W systems) among the lurker candidates.  
These appear to be WZ Sge stars, which 
often have outburst intervals of a decade or more.
It would not be surprising if {\it all} of DN-W systems
in Table \ref{tab:lurkers} were to outburst
within the next couple of decades, but it remains
possible that their outburst intervals could stretch
still longer.

\section{Acknowledgments}

{The National Science Foundation supported portions of 
this work through grants AST-0307413, AST-0708810, and
AST-1008217. 

We thank Taichi Kato for useful discussions, and for his tireless
efforts compiling reliable superhump parameters, and we are grateful
to the operators of the vsnet-alert system at Kyoto University.
This work made extensive use of the online archive of the 
vsnet-alert messages, which retain their value long after the
outbursts are over, and we are especially grateful to those who
maintain it.  We also thank
Elm\'e Breedt for communicating the periods of two of
the SDSS non-outbursters, and Tom Maccarone for pointing
out that SDSS might have missed some CVs harboring cooler
white dwarfs.

This research made use of the Digital 
Access to a Sky Century @ Harvard (DASCH), AAVSO, and Catalina Real-time 
Transient Survey databases. 
We acknowledge ESA Gaia, DPAC and the Photometric Science Alerts 
Team (http://gsaweb.ast.cam.ac.uk/alerts).
This research made use of the NASA/IPAC Infrared Science Archive, which is 
funded by NASA and operated by the California Institute of Technology. 
The DASCH project at Harvard is grateful for partial support from NSF grants 
AST-0407380, AST-0909073, and AST-1313370. }

Facilities: \facility{Hiltner, McGraw-Hill}

\object{FL Psc}
\object{CRTS CSS121120 J020633+205707}
\object{WY Tri}
\object{BB Ari}
\object{SDSSJ032015.29+441059.2}
\object{MASTER OT J034045.31+471632.2}
\object{V1024 Per}
\object{V1389 Tau}
\object{2E 0431.7+1756}
\object{V1208 Tau}
\object[ASAS-SN 15pq]{ }
\object{NAME Var Cam 06}
\object{SDSSJ075117.00+100016.2}
\object{SBSS 0755+600}
\object{TT Boo}
\object{QZ Lib }
\object[CSS170517:155156+145333]{ }
\object{1RXS J155720.3+180715}
\object{IX Dra}
\object{ASASSN -13at}
\object{V368 Peg}
\object{2MASS J18583091+4914326}
\object{RX J1946.2-0444}
\object{V1316 Cyg}
\object{1RXS J204455.9-115151}
\object{V444 Peg}
\object{ASASSN -14cl}
\object{V368 Peg}
\object{1RXS J230538.2+652155}
\object{CRTS J233848.7+281955}

\clearpage

\begin{deluxetable}{lrrr}
\tablecolumns{4}
\tablewidth{0pt}
\tablecaption{\label{tab:velocities} Radial Velocities}
\tablehead{
\colhead{Star} &
\colhead{Time\tablenotemark{a}} &
\colhead{$v_{\rm emn}$} &
\colhead{$\sigma$\tablenotemark{b}} \\
\colhead{} &
\colhead{} &
\colhead{[km s$^{-1}$]} &
\colhead{[km s$^{-1}$]} \\
}
\startdata
FL Psc   &    53327.7397  &   $-$7  &  19 \\
FL Psc   &    53327.7607  &  $-$62  &   7 \\
FL Psc   &    53327.7708  &   $-$9  &   8 \\
FL Psc   &    53327.7783  &    8  &   9 \\
FL Psc   &    53328.5730  &    9  &   6 \\
\enddata 
\tablenotetext{a}{Julian Date of mid-exposure minus 2,400,000, corrected to time
of arrival at the solar system barycenter.  The time system is UTC.}
\tablenotetext{b}{Uncertainty is derived from estimated counting-statistics
uncertainties.}
\tablecomments{Table \ref{tab:velocities} is published in its entirety in the electronic 
edition of The Astronomical Journal, A portion is shown here for guidance regarding its form and content.}
\end{deluxetable}

\clearpage

\begin{deluxetable}{lllrrcc}
\tablecolumns{7}
\footnotesize
\tablewidth{0pt}
\tablecaption{\label{tab:parameters} Fits to Radial Velocities}
\tablehead{
\colhead{Data set} & 
\colhead{$T_0$\tablenotemark{a}} & 
\colhead{$P$} &
\colhead{$K$} & 
\colhead{$\gamma$} & 
\colhead{$N$} &
\colhead{$\sigma$\tablenotemark{b}}  \\ 
\colhead{} & 
\colhead{} &
\colhead{(d)} & 
\colhead{(km s$^{-1}$)} &
\colhead{(km s$^{-1}$)} & 
\colhead{} &
\colhead{(km s$^{-1}$)} \\
}
\startdata
FL Psc & 53329.6182(10) & 0.05604(9) &  36(4) & $-17(3)$ & 36 &  13 \\ 
1RXS J0127+3808 & 53980.7751(10) & 0.06071(5)\tablenotemark{c} &  69(7) & $-48(5)$ & 95 &  23 \\ 
CSS 0206+20 & 58437.8125(16) & 0.06489(18)\tablenotemark{c} &  44(6) & $-28(5)$ & 53 &  18 \\
WY Tri  & 56943.7952(17) & 0.07569(7) &  108(15) & $-97(11)$ & 32 &  40 \\
BB Ari  & 54477.6396(12) & 0.07024722(11)\tablenotemark{d} &  79(9) & $ 9(6)$ & 62 &  29 \\ 
SDSS J0320+44 & 55514.883(2) & 0.06870(12) &  72(14) & $-28(10)$ & 58 &  32 \\ 
OT J0340+47 &  57320.696(3) & 0.0770(4)\tablenotemark{f} &  72(14) & $ 3(10)$ & 22 &  35 \\ 
V1024 Per & 55945.755(3) & 0.0706(3) &  23(6) & $10(4)$ & 41 &  14 \\ 
V1389 Tau & 55132.0190(18) & 0.0780(2)\tablenotemark{c} &  35(5) & $-58(4)$ & 24 &  13 \\ 
Gaia 19emm &  58865.5972(15) & 0.08580(6) &  81(10) & $ 87(7)$ & 46 &  31 \\
V1208 Tau & 53022.598(4) & 0.06813(15) &  50(17) & $-15(12)$ & 57 &  39 \\ 
ASAS-SN 15pq & 58867.6052(12) & 0.0800(2) &  92(10) & $ 8(6)$ & 20 &  23 \\
1RXS J05573+685 & 54124.976(4) & 0.0523(3) & 50(23) & $-52(16)$ & 56 &  37 \\ 
SDSS J0751+10 & 58865.6883(11) & 0.05922(5) &  58(7) & $ 90(5)$ & 45 &  22 \\ 
SBSS 0755+600 & 57435.0023(16) & 0.07334(7) &  99(14) & $-37(10)$ & 30 &  33 \\
TT Boo & 52437.751(3) & 0.07593962(7)\tablenotemark{e} &  31(7) & $-40(5)$ & 127 &  21 \\
QZ Lib & 53180.7919(15) & 0.06411(7)\tablenotemark{f} &  28(4) & $-44(3)$ & 48 &  14 \\
CSS 1551+14 & 57925.908(2) & 0.06966(16) &  53(12) & $-6(8)$ & 35 &  28 \\ 
SDSS J1557+18 & 57145.976(2) & 0.0810(2)\tablenotemark{c} &  66(12) & $-55(9)$ & 75 &  39 \\  
IX Dra & 56830.7477(20) & 0.06480(16) &  69(13) & $ 17(9)$ & 29 &  32 \\
ASAS-SN 13at & 57558.943(2) & 0.0792(2)\tablenotemark{c} &  33(6) & $-22(4)$ & 63 &  19 \\ 
KIC 11390659 & 58249.7782(15) & 0.07694(12)\tablenotemark{f} &  47(6) & $-48(4)$ & 57 &  18 \\ 
RX J1946.2-0444 & 54282.9008(13) & 0.074624(6) &  64(6) & $-1(5)$ & 62 &  22 \\
V1316 Cyg & 54646.877(3) & 0.07412(13) &  60(14) & $ 17(10)$ & 53 &  33 \\ 
ASAS-SN 14ds & 58363.704(3) & 0.1058(3) &  48(8) & $-35(5)$ & 55 &  25 \\ 
V444 Peg & 55070.649(3) & 0.09260(18) &  76(15) & $-96(10)$ & 47 &  34 \\ 
ASAS-SN 14cl & 56946.613(2) & 0.05859(4) &  26(5) & $ 25(4)$ & 36 &  13 \\
V368 Peg & 52577.888(3) & 0.0685(3) &  22(5) & $-5(4)$ & 45 &  13 \\ 
IPHAS 2305 & 54711.979(4) & 0.0702(3)\tablenotemark{f} &  27(8) & $-37(6)$ & 37 &  20 \\ 
NSV14652 & 56945.6030(14) & 0.07824(8) &  81(11) & $-24(7)$ & 26 &  25 \\
\enddata

\tablecomments{Parameters of least-squares sinusoid fits to the radial
velocities, of the form $v(t) = \gamma + K \sin(2 \pi(t - T_0)/P$.}
\tablenotetext{a}{Heliocentric Julian Date minus 2400000.  The epoch is chosen
to be near the center of the time interval covered by the data, and
within one cycle of an actual observation.}
\tablenotetext{b}{RMS residual of the fit.}
\tablenotetext{c}{Becaues the cycle count between runs is unknown, the 
period given here is the weighted average of periods derived from
the individual observing runs.  The remaining parameters are from a fit
to the entire data set.}
\tablenotetext{d}{The precise period given reflects a likely choice of cycle count between
observing runs; see discussion in text.}
\tablenotetext{e}{The period given is the most likely precise period, but other choices
of cycle count between runs have similar likelihood and give slightly different periods;
see text.}
\tablenotetext{f}{Daily cycle count inferred using the known superhump period.}
\end{deluxetable}

\begin{deluxetable}{lrrrrrrl}
\tablecolumns{7} 
\tablecaption{\label{tab:bouncers} Candidate Period Bouncers}
\tablehead{
\colhead{Name} &
\colhead{$\alpha$\tablenotemark{a}} &
\colhead{$\delta$} &
\colhead{$\epsilon$} &
\colhead{$q$} &
\colhead{$P_{\rm orb}$} &
\colhead{$P_{\rm orb}$} \cr
\colhead{Ref.\tablenotemark{b}} &
\colhead{[h:m:s]} &
\colhead{[d:$'$:$''$]} &
\colhead{} &
\colhead{} &
\colhead{[d]} &
\colhead{[min]} &
\colhead{} \\ 
}
\startdata
OT J111218-353837    &  11:12:17.40 &  $-$35:38:29.0 & 0.0085 & 0.0439 & 0.05847 &    84.2 & 1 \\
EZ    Lyn            &  08:04:34.20 &  +51:03:49.2 & 0.0103 & 0.0526 & 0.05901 &    85.0 & 1,3 \\
PNV 17144255-2943481 &  17:14:42.60 &  $-$29:43:45.0 & 0.0090 & 0.0464 & 0.05956 &    85.8 & 7 \\
MASTER 211258.65+242145.4 &  21:12:58.65 &  +24:21:45.4 & 0.0083 & 0.0431 & 0.05973 &    86.0 & 5 \\
ASASSN-14cv          &  17:43:48.58 &  +52:03:46.8 & 0.0083 & 0.0430 & 0.05992 &    86.3 & 7 \\
EG    Cnc            &  08:43:04.04 &  +27:51:49.9 & 0.0061 & 0.0323 & 0.05997 &    86.4 & 1 \\
ASASSN-16js          &  00:51:19.07 &  $-$65:57:16.9 & 0.0098 & 0.0506 & 0.06034 &    86.9 & 9 \\
ASASSN-17fn          &  10:35:28.34 &  +54:19:07.5 & 0.0102 & 0.0524 & 0.06096 &    87.8 & 10 \\
OT J220559.40-341434.9 &  22:05:59.40 &  $-$34:14:34.9 & 0.0116 & 0.0590 & 0.06129 &    88.3 & 9 \\
ASASSN-15na          &  19:19:08.84 &  $-$49:45:41.0 & 0.0119 & 0.0603 & 0.06297 &    90.7 & 8 \\
QZ Lib               &  15:36:16.02 &  $-$08:39:08.6 & 0.0188 & 0.0401 & 0.06411 &    92.3 & T \\
CSS160414:122625-113303 &  12:26:25.43 &  $-$11:33:03.3 & 0.0048 & 0.0258 & 0.06420 &    92.4 & 9 \\
\enddata
\tablenotetext{a}{Coordinates are for J2000 and are given for identification.}
\tablenotetext{b}{References: T = this work, also \citet{pala18}; 1 = \citet{kato1}; 3 = \citet{kato3},
5 = \citet{kato5}; 7 = \citet{kato7}; 8 = \citet{kato8}; 9 = \citet{kato9}; 10 = \citet{kato10}.}
\end{deluxetable}

\begin{deluxetable}{lrlrrrrrrr}
\tablecolumns{10}
\tablewidth{0pt}
\tablecaption{\label{tab:lurkers} SDSS CVs Without Known Outbursts}
\tablehead{
\colhead{Object} &
\colhead{Ref. \tablenotemark{a}} &
\colhead{Type} & 
\colhead{$P_{\rm orb}$} &
\colhead{ref.\tablenotemark{b}} & 
\colhead{$g$} &
\colhead{$G$ \tablenotemark{c}} & 
\colhead{$\pi$} & 
\colhead{$\pi$ error} &
\colhead{PM} \\
\colhead{SDSS} &
\colhead{} &
\colhead{} &
\colhead{[min]} &
\colhead{} &
\colhead{[mag.]} &
\colhead{[mas]} &
\colhead{[mas]} &
\colhead{[mas]} &
\colhead{[mas yr$^{-1}$]} 
}
\startdata
SDSS J003941.06+005427.5  & 4   & DN-W   &  91 & 1   & 20.57  & 20.89 & $-$1.3322 & 1.9534 &  21.35  \cr
SDSS J004335.14$-$003729.8  & 3   & DN-W   &  82 & 2   & 19.84  & 19.88 & 2.9876 & 0.5716 &  29.45  \cr
SDSS J023003.79+260440.3  & 7   & DN-2   & \nodata & \nodata   & 19.91  & 19.05 & 1.6944 & 0.3488 &   6.80  \cr
SDSS J080534.49+072029.1  & 6   & DN-2   & 330 & 3   & 18.52  & 17.90 & 0.5133 & 0.1708 &   6.04  \cr
SDSS J083754.64+564506.7  & 6   & DN   & \nodata & \nodata   & 18.97  & \nodata & \nodata & \nodata & \nodata \cr
SDSS J085623.00+310834.1  & 4   & DN-W   & \nodata & \nodata   & 19.99  & 20.21 & 1.4812 & 0.9352 &  11.30  \cr
SDSS J090403.48+035501.2  & 3   & DN-W   &  86 & 4   & 19.24  & 19.32 & 3.7443 & 0.6808 &   4.61  \cr
SDSS J090452.09+440255.4  & 3   & DN-W   & \nodata & \nodata   & 19.38  & 19.46 & 2.7236 & 0.4266 &  16.76  \cr
SDSS J101037.05+024915.0  & 2   & DN   & 138 & 5  & 20.76  & \nodata & \nodata & \nodata & \nodata \cr
SDSS J110706.76+340526.8  & 6   & DN   & \nodata & \nodata   & 19.48  & 18.44 & 0.6122 & 0.2722 &   5.30  \cr
SDSS J121607.03+052013.9  & 3   & DN   &  99 & 6   & 20.12  & 20.20 & 2.6277 & 0.9565 &  67.91  \cr
SDSS J121913.04+204938.3  & 8   & DN-W   & \nodata & \nodata   & 19.17  & 19.24 & 3.5769 & 0.3884 &  71.86  \cr
SDSS J125641.29$-$015852.0  & 1   & DN-W   & 103/111\tablenotemark{d} & 5 & 20.12  & 20.59 & 4.0089 & 1.0855 &  16.19  \cr
SDSS J125834.77+663551.6  & 2   & DN   & \nodata & \nodata   & 20.20  & 19.82 & 0.3555 & 0.3570 &   5.96  \cr
SDSS J143317.78+101123.3  & 6   & DN-W   &  78 & 7   & 18.55  & 18.59 & 4.4539 & 0.2027 &  54.94  \cr
SDSS J145003.12+584501.9  & 2   & DN   & \nodata & \nodata   & 20.64  & \nodata & \nodata & \nodata & \nodata \cr
SDSS J151413.72+454911.9  & 4   & DN-W   & \nodata & \nodata   & 19.68  & 19.71 & 3.2272 & 0.3326 &  78.88  \cr
SDSS J155531.99$-$001055.0  & 1   & DN   & 114 & 8   & 19.36  & 18.99 & 1.5963 & 0.2991 &  10.21  \cr
SDSS J171145.08+301320.0  & 3   & DN-W   &  80 & 9   & 20.25  & 20.21 & 2.8630 & 0.5112 &  15.26  \cr
SDSS J171247.71+604603.3  & 1   & DN-2   & \nodata & \nodata   & 19.95  & 18.80 & 1.2925 & 0.1694 &  11.20  \cr
SDSS J172601.96+543230.7  & 1   & DN   & \nodata & \nodata   & 20.52  & \nodata & \nodata & \nodata & \nodata \cr
SDSS J202520.13+762222.4  & 5   & DN-2   & \nodata & \nodata   & 21.83  & 20.56 & $-$0.3667 & 1.2581 &   6.01  \cr
\enddata
\tablenotetext{a}{Paper in which the object was identified; 1 through 8 
respectively refer to \citet{szkodyi,szkodyii,szkodyiii,szkodyiv,
szkodyv,szkodyvi,szkodyvii} and \citet{szkodyviii}.}
\tablenotetext{b}{References for periods: 
1 : \citet{southworth10},
2 : \citet{southworth08},
3 : \citet{thorsdss},
4 : \citet{woudt12},
5 : E. Breedt, private communication, from VLT/FORS velocities.
6 : \citet{southworth06},
7 : \citet{littlefair08},
8 : \citet{southworth07},
9 : \citet{dillon08}}

\tablenotetext{c}{$G$ magnitudes and astrometric parameters are from the 
Gaia Data Release 2.}

\tablenotetext{d}{Two alias periods are possible.}

\end{deluxetable}

\begin{deluxetable}{rcccc}
\tablecolumns{5}
\tablecaption{\label{tab:lurkertype} SDSS CVs with Dwarf Nova Spectra}
\tablehead{
\colhead{} &
\colhead{DN} & 
\colhead{DN\_2} & 
\colhead{DW\_W} &
\colhead{Totals} 
}
\startdata 
Prev. known: & 29 & 5 & 3 & 37 \cr
O/B found:  & 81 & 13 & 26 & 120 \cr
O/B not found: & 8 & 4 & 10 & 22 \cr
Totals:    & 118 & 22  & 39 & 179 \cr
\enddata
\end{deluxetable}

\end{document}